\documentclass[]{aastex631}

 \usepackage{amsmath}
\usepackage{multirow}
\usepackage{makecell}
\usepackage{soul}
\def\cof {$^{12}\mathrm{CO}~(J=1\rightarrow0)$}
\def\cos {$^{13}\mathrm{CO}~(J=1\rightarrow0)$}

\def\largestQ1{G034.4$+$01.0}
\def\cot {$\mathrm{C}^{18}\mathrm{O}~(J=1\rightarrow0)$}

\def\cofs {$^{12}\mathrm{CO}$}
\def\coss {$^{13}\mathrm{CO}$}
\def\cots{$\mathrm{C}^{18}\mathrm{O}$}
\def\deg  {\ifmmode {^\circ}\else {$^\circ$}\fi}

\def\kms     {km~s$^{-1}$}
 \newcommand{\tbreak}{$T_{\rm break}$}  
 \newcommand{\tmean}{$\langle T\rangle$}

\usepackage{booktabs}
\usepackage{chngcntr}

\date{\today}
\submitjournal{ApJL}

\shorttitle{Flux-Intensity Relation}
\shortauthors{Yan et al.} 

\begin{document}




\title{On the Flux-Intensity Relation of Molecular Clouds} 

\correspondingauthor{Ji Yang}
\email{jiyang@pmo.ac.cn, qzyan@pmo.ac.cn}

\author[0000-0003-4586-7751]{Qing-Zeng Yan}
\affil{Purple Mountain Observatory and Key Laboratory of Radio Astronomy,\\
 Chinese Academy of Sciences, 10 Yuanhua Road, Qixia District, Nanjing 210033, People's Republic of China}

\author[0000-0001-7768-7320]{Ji Yang}
\affil{Purple Mountain Observatory and Key Laboratory of Radio Astronomy,\\
 Chinese Academy of Sciences, 10 Yuanhua Road, Qixia District, Nanjing 210033, People's Republic of China}

 \author[0000-0002-0197-470X]{Yang Su }
 \affil{Purple Mountain Observatory and Key Laboratory of Radio Astronomy,\\
 Chinese Academy of Sciences, 10 Yuanhua Road, Qixia District, Nanjing 210033, People's Republic of China}

 \author[0000-0002-3904-1622]{Yan Sun}
\affil{Purple Mountain Observatory and Key Laboratory of Radio Astronomy,\\
 Chinese Academy of Sciences, 10 Yuanhua Road, Qixia District, Nanjing 210033, People's Republic of China}

 \author[0000-0003-2549-7247]{Shaobo Zhang}
\affil{Purple Mountain Observatory and Key Laboratory of Radio Astronomy,\\
 Chinese Academy of Sciences, 10 Yuanhua Road, Qixia District, Nanjing 210033, People's Republic of China}
 
 \author[0000-0003-2418-3350]{Xin Zhou}
\affil{Purple Mountain Observatory and Key Laboratory of Radio Astronomy,\\
 Chinese Academy of Sciences, 10 Yuanhua Road, Qixia District, Nanjing 210033, People's Republic of China}

\author[0000-0001-8923-7757]{Chen Wang}
\affil{Purple Mountain Observatory and Key Laboratory of Radio Astronomy,\\
 Chinese Academy of Sciences, 10 Yuanhua Road, Qixia District, Nanjing 210033, People's Republic of China}

\author[0000-0003-3139-2724]{Yiping Ao}
\affil{Purple Mountain Observatory and Key Laboratory of Radio Astronomy,\\
 Chinese Academy of Sciences, 10 Yuanhua Road, Qixia District, Nanjing 210033, People's Republic of China}

\author[0000-0003-3151-8964]{Xuepeng Chen}
\affil{Purple Mountain Observatory and Key Laboratory of Radio Astronomy,\\
 Chinese Academy of Sciences, 10 Yuanhua Road, Qixia District, Nanjing 210033, People's Republic of China}

\author[0000-0003-4767-6668]{Min Wang}
\affil{Purple Mountain Observatory and Key Laboratory of Radio Astronomy,\\
 Chinese Academy of Sciences, 10 Yuanhua Road, Qixia District, Nanjing 210033, People's Republic of China}






\begin{abstract}  
In this work, we report a study on the relationship between flux and intensity for molecular clouds. Our analysis is established on high-quality CO images from the Milky Way Imaging Scroll Painting (MWISP) project.  The new flux-intensity relation characterizes the flux variation of molecular clouds above specific intensity levels. We found that the flux-intensity relation exhibits two prominent features.  First, the flux-intensity relation generally follows exponential shapes; secondly, hierarchical structures of molecular clouds are imprinted on flux-intensity relations. Specifically, \cofs\ flux-intensity relations are composed of one or more exponential segments, and for molecular clouds with segmented flux-intensity relations, the edge and the flux of the  high-temperature component are strikingly consistent with \coss\ emission. Further analysis shows that a similar relationship also exists between \coss\ flux-intensity relations and \cots\ emission. The mean brightness temperature of molecular clouds is tightly associated with the decay rate of flux, the break temperature of exponential segments, and, to a certain extent, the flux fraction of the high-temperature component. Broadly, the flux-intensity relation of a molecular tracer, either in optically thick or in optically thin cases, has the capability to outline the silhouette of internal structures of molecular clouds, proving to be a potent tool for probing structures of molecular clouds. 
\end{abstract}

\keywords{Molecular clouds (1072);  Molecular spectroscopy (2095);  Observational astronomy (1145); Astrostatistics techniques (1886); Astronomy data analysis (1858)}


\section{Introduction} \label{sec:intro}

Structures of molecular clouds are usually portrayed through 3D geometric coordinates, namely analytic studies \citep[e.g.,][]{2001Natur.409..159A,2001ApJ...546..980O,2021ApJ...919...35Z}. To simplify calculations or due to deficient observations, it is frequently analyzed as a function of radial distance ($r$) relative to a specific location in 1D cases. Those studies explore various aspects of molecular clouds, with a particular emphasis on the volume density and temperature structures. For example, from low- to high-volume density regions, practically used tracers are dust extinction maps \citep{1985A&A...149..273C,2001Natur.409..159A}, sub-millimeter continuum \citep{1994MNRAS.268..276W}, and molecular spectra \citep{1985ApJ...297..436A,2016ApJ...822...10K}. The analysis of density or temperature radial profiles are also applicable to filamentary structures interior to molecular clouds \citep{2011A&A...529L...6A,2011A&A...533A..34H,2013A&A...550A..38P}.

Parallel to the analytic studies, an alternative approach, the statistical study, utilizes statistical tools to explore molecular cloud properties. This statistical approach ignores the position pattern and concentrates on the statistics of cloud properties. For instance, statistical analyses have already been used to study the fractal dimension \citep[e.g.,][]{1991ApJ...378..186F,1996ApJ...471..816E}, structure functions based on intensity maps \citep[e.g.,][]{1998A&A...336..697S,2003ApJ...583..308P}, statistics on three-dimensional velocity field \citep{2006A&A...452..223O}, turbulent velocities \citep[e.g.,][]{2004ApJ...615L..45H,2012A&ARv..20...55H,2015ARA&A..53..583H}, column density probability distribution functions (N-PDFs) \citep{2008A&A...489..143L,2009ApJ...692...91G,2009A&A...508L..35K,2010MNRAS.408.1089T,2013ApJ...766L..17S,2014Sci...344..183K,2014A&A...565A..18A}, histogram of oriented gradients (HOG) \citep{2019A&A...622A.166S}, the spectra correlation function (SCF) \citep{2003ApJ...588..881P}, power spectra \citep{2018ApJ...856..136P}, and the width-size relation \citep{1998ApJ...504..223G}. These researches primarily delve into turbulence, as well as the formation and evolution mechanism of molecular clouds. However, statistics on basic observational properties, such as flux, are rarely investigated.

 Analytic studies and statistical studies are not independent but rather complementary to each other. To investigate the internal structures of molecular clouds, we need both relative positions and property statistics of those structures. For example, by comparing the spatial and velocity relationships between cores and filaments, along with property distributions of filaments, \citet{2013A&A...554A..55H} concluded that hierarchical fragmentation proceeds core formation. In fact, investigations of hierarchical structures of molecular clouds typically include decomposition based on positions, as well as analyses involving statistics \citep{2008ApJ...679.1338R,2015MNRAS.454.2067C}.

In this work, we take the statistical approach to investigate the flux-intensity relation of individual molecular clouds and its behaviors in a cloud sample.  The content of this work is organized as follows. In \S\ref{sec:codata}, we briefly describe the CO data and the method to construct CO samples. The flux-intensity relation measurement methodology and results are demonstrated in \S\ref{sec:results}. We discuss more details on the flux-intensity relation in \S\ref{sec:dis} and summarize conclusions in \S\ref{sec:conclusion}.

\section{CO Data and Cloud Samples}
\label{sec:codata}

The CO data are part of the MWISP survey \citep{2014AJ....147...46Z,2016A&A...588A.104G,2019ApJS..240....9S}. The MWISP survey is carried out with the PMO-13.7m millimeter telescope, which is armed with a 3$\times$3 multi-beam sideband-separating Superconducting SpectroScopic Array Receiver (SSAR) \citep{2012ITTST...2..593S}. The half-power beam width (HPBW) of the PMO-13.7m telescope is about 50\arcsec\ at 115 GHz.

The MWISP survey simultaneously observes three CO isotopologue lines, \cof, \cos, and \cot, at a velocity resolution of $\sim$0.16 \kms, and the root mean square (rms) noise is measured at about 0.5 K, 0.3 K, and 0.3 K, respectively. For further details on instruments and observation strategies, we refer readers to \citet{2012ITTST...2..593S} and \citet{2019ApJS..240....9S}. 

In this work, we construct molecular cloud samples from the whole Phase I data cubes of the MWISP survey. The PPV range of data cubes is $9.75\deg\leq l \leq229.25\deg$, $|b|\leq5.25\deg$, and $\rm -200 \leq \mathit{V}_{\rm LSR}\leq 300\ km\ s^{-1}$. The algorithm employed to build molecular cloud samples is DBSCAN \citep{Ester:1996:DAD:3001460.3001507,2020ApJ...898...80Y}. We run the DBSCAN algorithm on data cubes of three CO lines, and the boundary of a molecular cloud is defined by the closed surface of \cofs ~in PPV space.  The  \cofs ~boundary is more extended than those by \coss\ and \cots. As a consequence, \coss\ or \cots\  clumps within a \cofs ~boundary are regarded as internal structures of the cloud.

In this work, we focus on molecular clouds that have both \cofs\ and  \coss\ emission, and all these clouds are complete in PPV space.  We primarily use the flux-intensity relation of \cofs, whose emission is the most extended and whose signal-to-noise ratios (SNRs) are the highest. Parameters of five examples of molecular clouds are listed in Table \ref{Tab:cloudpara}. The flux of \cofs\ is assessed by DBSCAN, while the flux of \coss\ and \cots\ is measured by stacking over \cofs\ and \coss\ PPV envelopes (obtained with DBSCAN), respectively.

 Two molecular clouds are rejected due to inadequate data points ($\leq$ 5, see section \ref{sec:doubleexp}), and flux-intensity relations of the remaining 10866 clouds are measured. All these 10,866 molecular clouds have  significant \coss\ emission, and 305 of them have significant \cots\ emission. The online catalog of these clouds is available at ScienceDB (DOI:10.57760/sciencedb.17486).

 \begin{deluxetable}{ccccccccc}[h]
\tablecaption{PPV parameters of 10,866 molecular clouds\tablenotemark{e}. \label{Tab:cloudpara}}
 
\tablehead{
 \colhead{Index}  & \colhead{$l$\tablenotemark{a}} & \colhead{$b$\tablenotemark{a}} &  \colhead{$V_{\rm LSR}$\tablenotemark{a}}  & \colhead{Angular area} & \colhead{\cofs\ Flux} & \colhead{\coss\ Flux\tablenotemark{b}}  & \colhead{\cots\ Flux\tablenotemark{c}} & \tmean\tablenotemark{d} \\
\colhead{ }  &  \colhead{(deg)}  &  \colhead{(deg)}    & \colhead{(\kms)} & \colhead{(armin$^2$)} &    \colhead{(K \kms\ armin$^2$)} &    \colhead{(K \kms\ armin$^2$)} &    \colhead{(K \kms\ armin$^2$)} & \colhead{(K)}
} 
\startdata
1932340  & 105.292 & 4.465 & -82.49& 40.25 & 81.38 & 5.46 & -- & 1.3 \\
11061298 & 16.399 & 0.064 & 130.53 & 38.75  & 78.71 & 14.29 & -- & 1.7 \\
3725139  & 156.040 & -2.993 & -23.59 & 272.00 & 1339.66 & 142.02 & -- & 2.5 \\
6856826  & 64.707 & -1.680 & 18.13 & 64.25 & 151.03 & 23.27 & 2.16 & 2.5 \\
6366510 & 16.897 & -2.428 & 17.91& 1592.50   & 30464.66 & 6821.46 & 726.92 & 5.0 \\
6366510 & 16.897 & -2.428 & 17.91& 1592.50   & 30464.66 & 6821.46 & 726.92 & 5.0 \\
: & : & :  & :  &  :    &  :  &  :  &  :   &  :  \\
\enddata
\tablenotetext{a}{Average values weighted by intensity.}
\tablenotetext{b}{Measured by stacking \coss\ voxels over \cofs\ PPV envelopes of molecular clouds.}
\tablenotetext{c}{Measured by stacking \cots\ voxels over \coss\ PPV envelopes of molecular clouds.}
\tablenotetext{d}{Mean \cofs\ brightness temperature across all voxels within \cofs\ PPV envelopes of molecular clouds.}
\tablenotetext{e}{ A full version of this molecular cloud catalog can be accessed at DOI:10.57760/sciencedb.17486. }
\end{deluxetable}

\label{sec:sample}

\section{Results}

\label{sec:results}

The approach of this study is to establish the relationship between flux and intensity for individual molecular clouds and to examine their distribution characteristics. A molecular cloud usually exhibits hierarchical patterns and irregular morphology, and this irregularity brings uncertainties in the analysis of cloud structure regarding cloud radius or other quantities dependent on coordinates. To minimize uncertainties introduced by morphology,  we measure the flux of a cloud above certain intensity levels, i.e., the flux above specific isochrones.

\subsection{Single-Exponential Clouds}

 \label{sec:singleexp}
For an individual cloud,  its flux-intensity relation is established by measuring the \cofs\ flux above a certain brightness temperature contour $T$. In this work, the brightness temperature and intensity are used interchangeably, and we primarily use \cofs\ due to its high SNRs. 

Specifically, the intensity contour starts from the peak temperature to the 2$\sigma$ ($\sim$ 1 K) edge, with a step of 0.25 K. Fine steps tend to yield smooth flux-intensity relations but with low flux SNRs, while coarse steps may miss details of the flux-intensity relation. We find 0.5$\sigma$ (about 0.25 K) is an appropriate choice. For each intensity contour $T$, the corresponding flux is measured by summing across all voxels with intensity larger than $T$, and this measurement is restricted within the boundaries of molecular clouds defined by the \cofs\ emission.

As a result, we find that flux-intensity relations of many molecular clouds exhibit an excellent fit to an exponential function. The  exponential function is defined as 
\begin{equation}
\label{eq:exp}
F_{\rm CO}= a\exp(-\alpha T)+c, 
\end{equation} 
where $\alpha$ describes the steepness of the exponential function. $a$ and $c$ are the other two parameters, and values of $\alpha$, $a$, and $c$ vary from cloud to cloud.

\begin{figure}[h]
\gridline{\fig{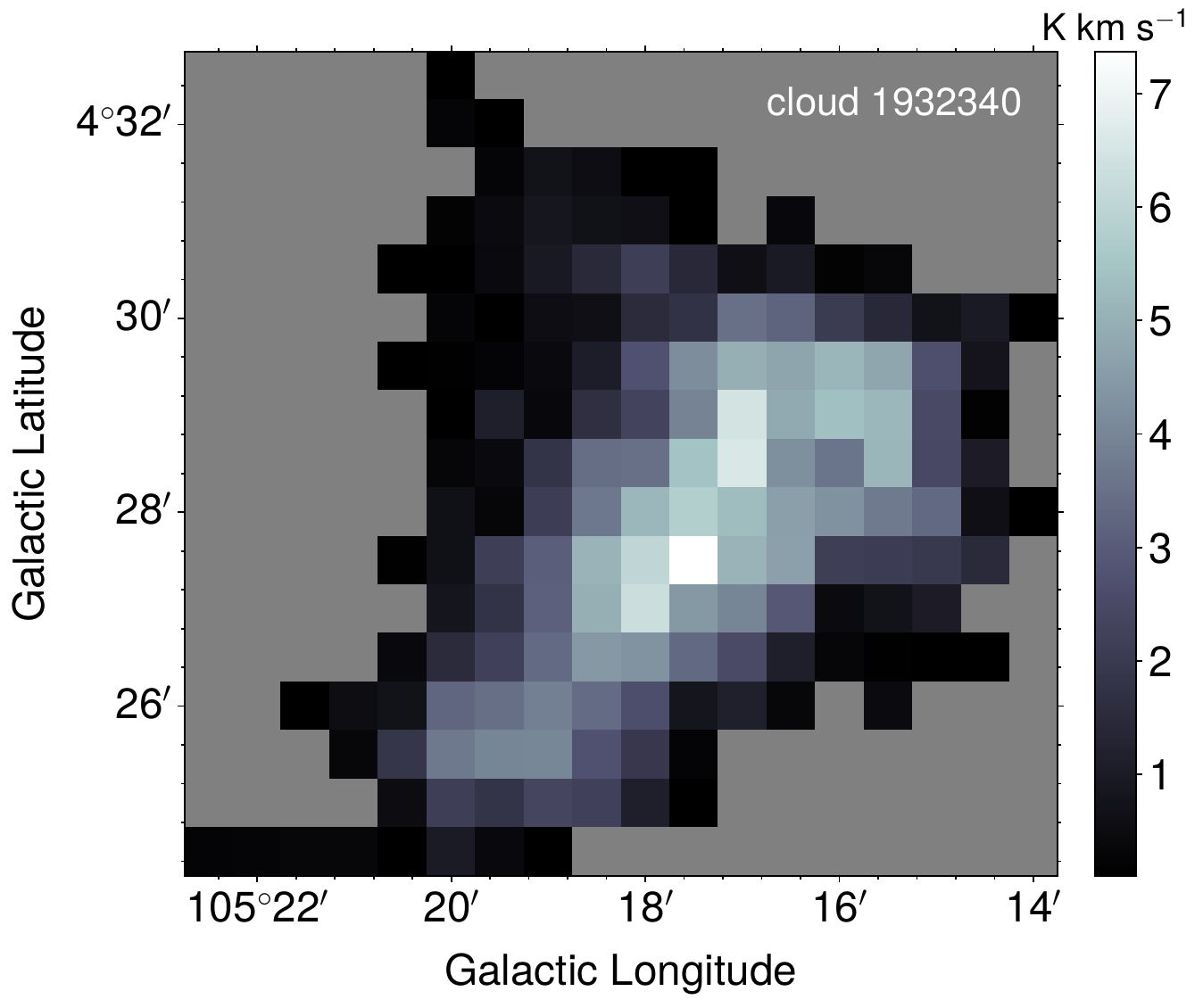}{0.45\textwidth}{(a) Image of cloud 1932340.}  \fig{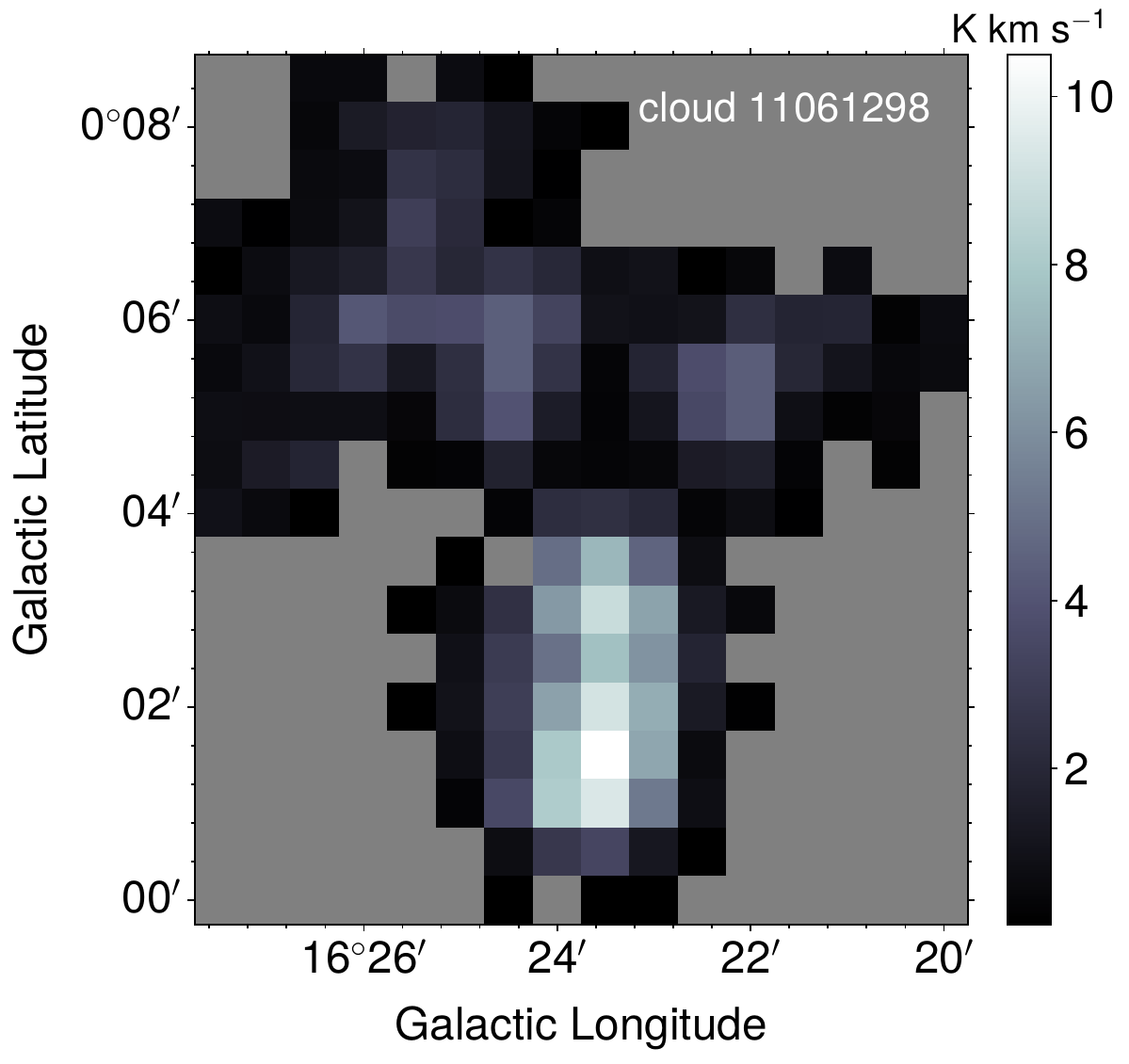}{0.45\textwidth}{(b) Image of cloud 11061298.}  } 
\gridline{\fig{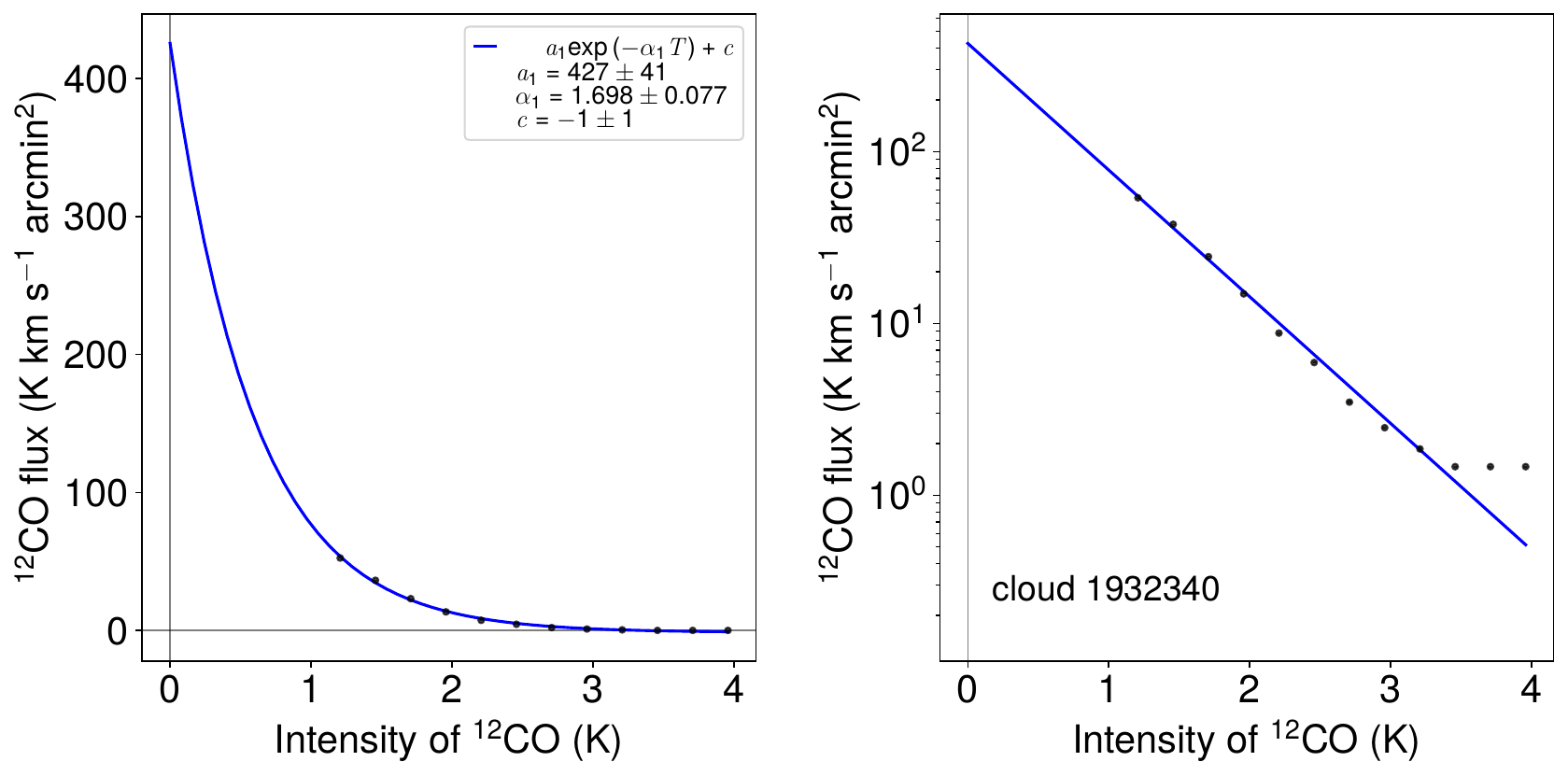}{0.45\textwidth}{(c) Flux-intensity relation of cloud 1932340.}  \fig{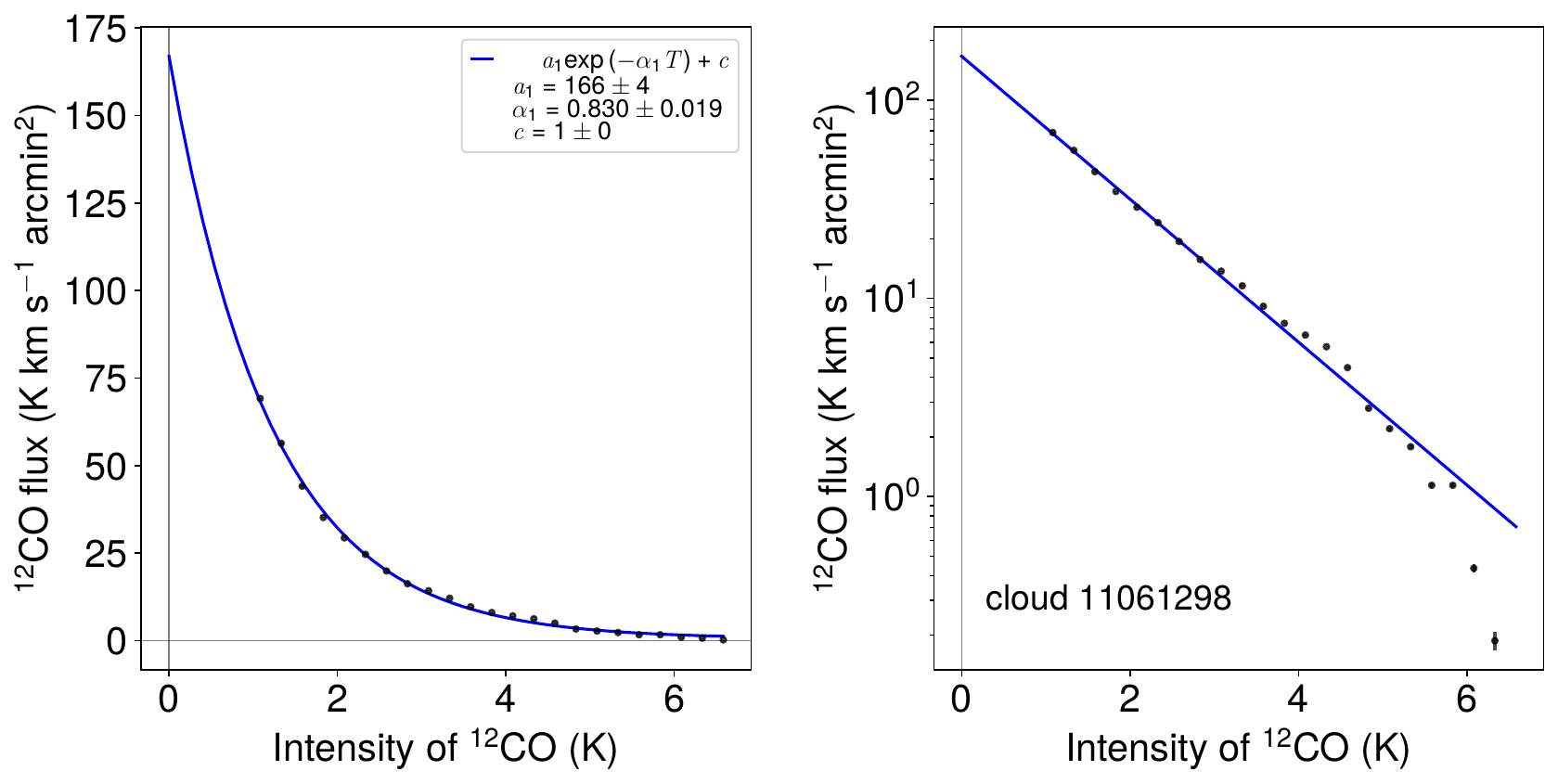}{0.45\textwidth}{(d) Flux-intensity relation of cloud 11061298.}  } 
\caption{Integrated \cofs\ intensity images and flux-intensity relations of two single-exponential molecular clouds. See Table \ref{Tab:cloudpara} for detailed parameters of the two molecular clouds. For added clarity on exponential properties, plots in logarithmic scale are also displayed, where the constant $c$ in Equation \ref{eq:exp} has been  subtracted. \label{fig:singleexp}}  
\end{figure}

As examples, Figure \ref{fig:singleexp} shows the flux-intensity relations and the exponential fittings of two molecular clouds. In these cases, the fitting to the exponential form is remarkably good, especially considering the irregular morphologies and  complex textures of molecular clouds.

 In total, 499 molecular clouds exhibit single-exponential  flux-intensity relations. Additionally, we also examined  many molecular clouds whose \coss\ emission is weak, and their flux-intensity relations are also in exponential form. 

Fitting results show that values of $\alpha$ range from 0.01 to 1.96 K$^{-1}$. Details of the $\alpha$ distribution are demonstrated in section \ref{sec:samplingdis}.

\subsection{Double-Exponential Clouds}
 \label{sec:doubleexp}

In addition to the single-exponential flux-intensity relation mentioned above, we observed that substantially more molecular clouds follow segmented flux-intensity relations. A flux-intensity relation with two exponential segments is defined as 
 \begin{equation}
 \label{eq:exp2}
 F_{\rm CO}= 
\begin{cases}
a_1\exp(-\alpha_1T)+c &  T< T_{\rm break} \\
a_2\exp(-\alpha_2(T-T_{\rm break}))&   T \geq T_{\rm break},  \\
  \end{cases}
\end{equation} 
where $T_{\rm break}$ is the break temperature of this piecewise function, i.e., the joint temperature of two exponential segments. The constant term of the second segment  is dropped  for a better fitting of $T_{\rm break}$ (judged by eyes). It is possible that a third exponential segment exists, but fitting of two exponential components is mostly satisfied.

\begin{figure}[h]
\gridline{\fig{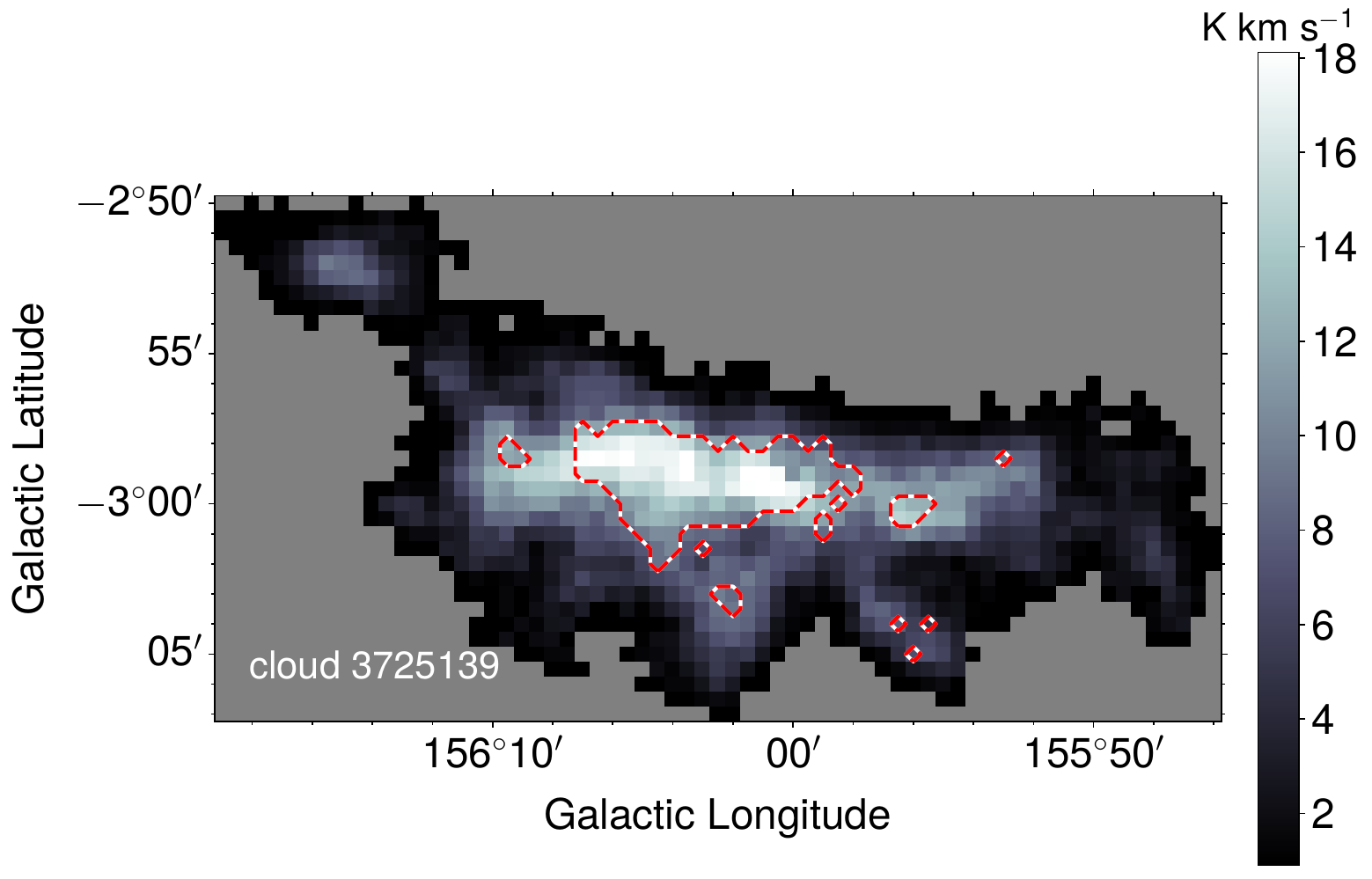}{0.375\textwidth}{(a) Image of cloud 3725139.}  \fig{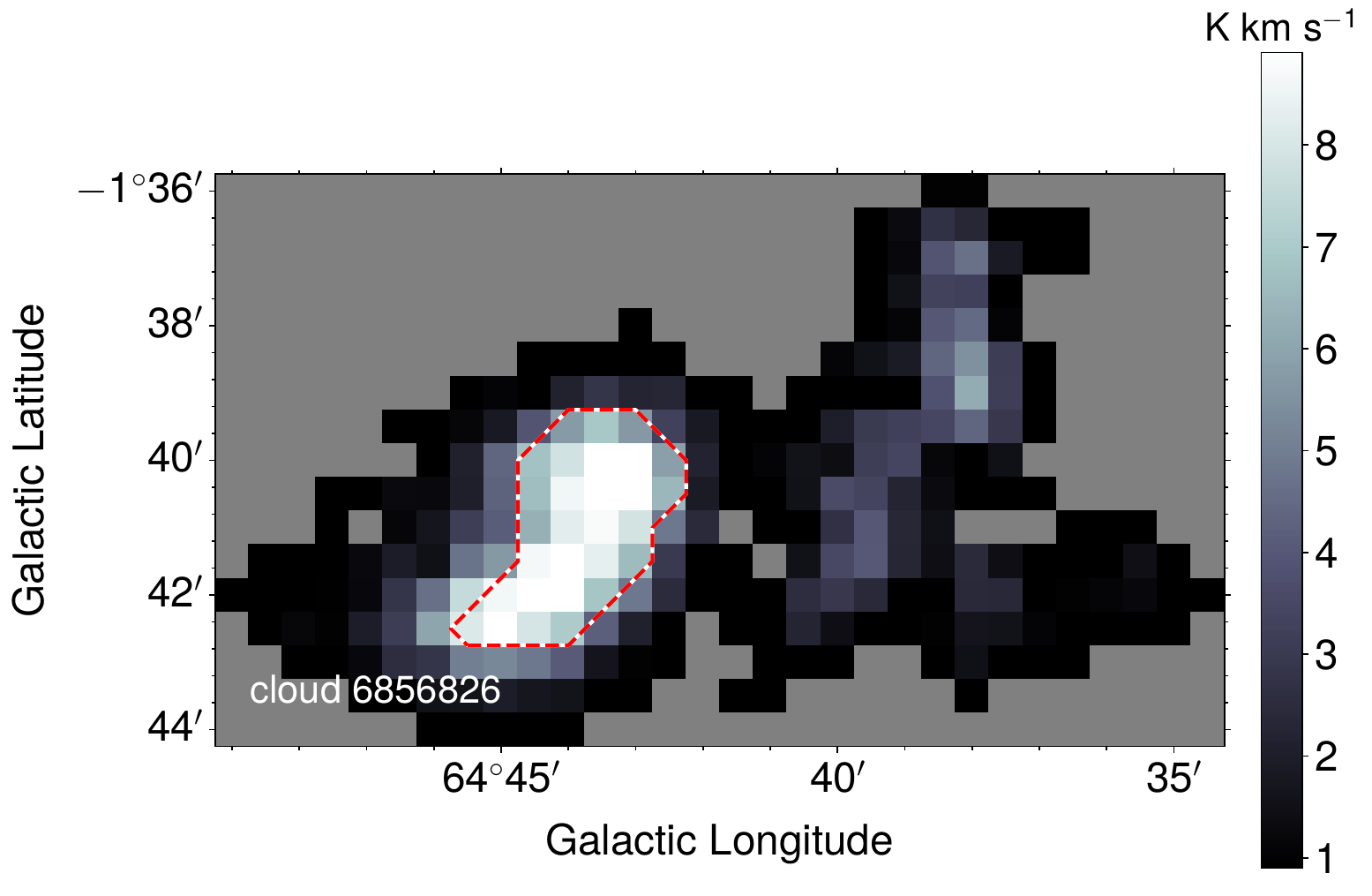}{0.375\textwidth}{(b) Image of cloud 6856826.}  \fig{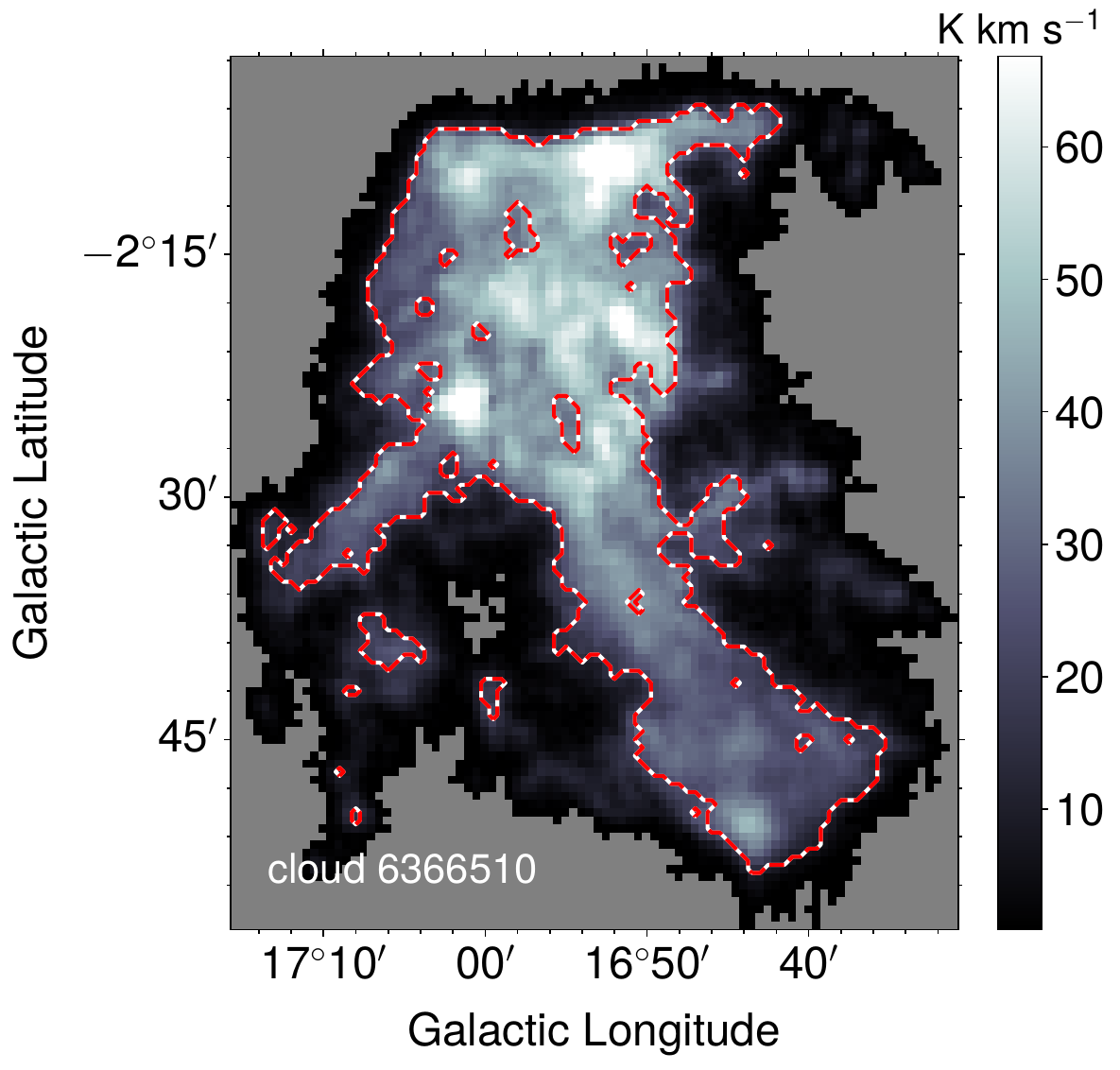}{0.25\textwidth}{(c) Image of cloud 6366510.} } 
\gridline{ \fig{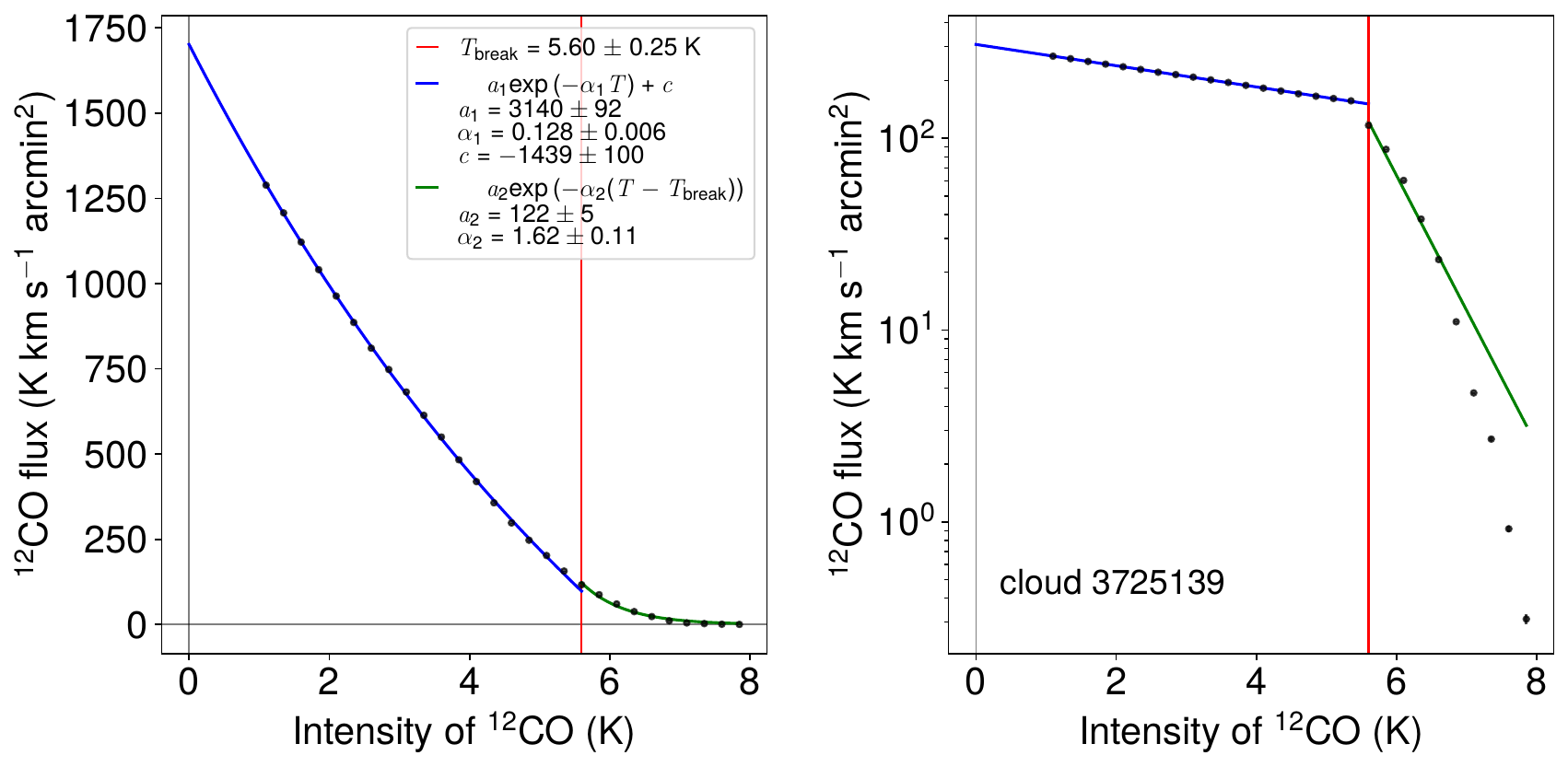}{0.33\textwidth}{(d) Flux-intensity relation of cloud 3725139.}  \fig{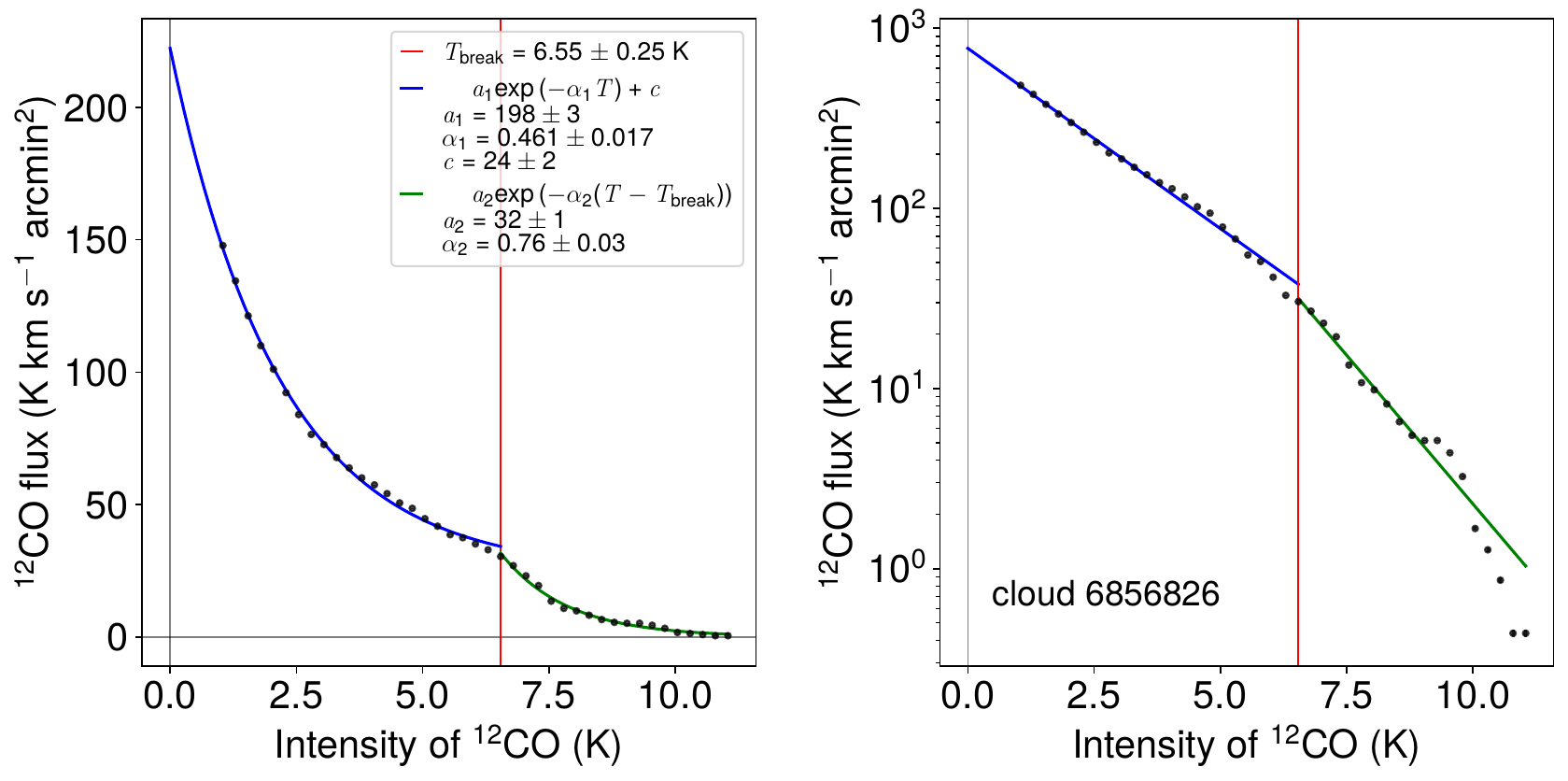}{0.33\textwidth}{(e) Flux-intensity relation of cloud 6856826.}  \fig{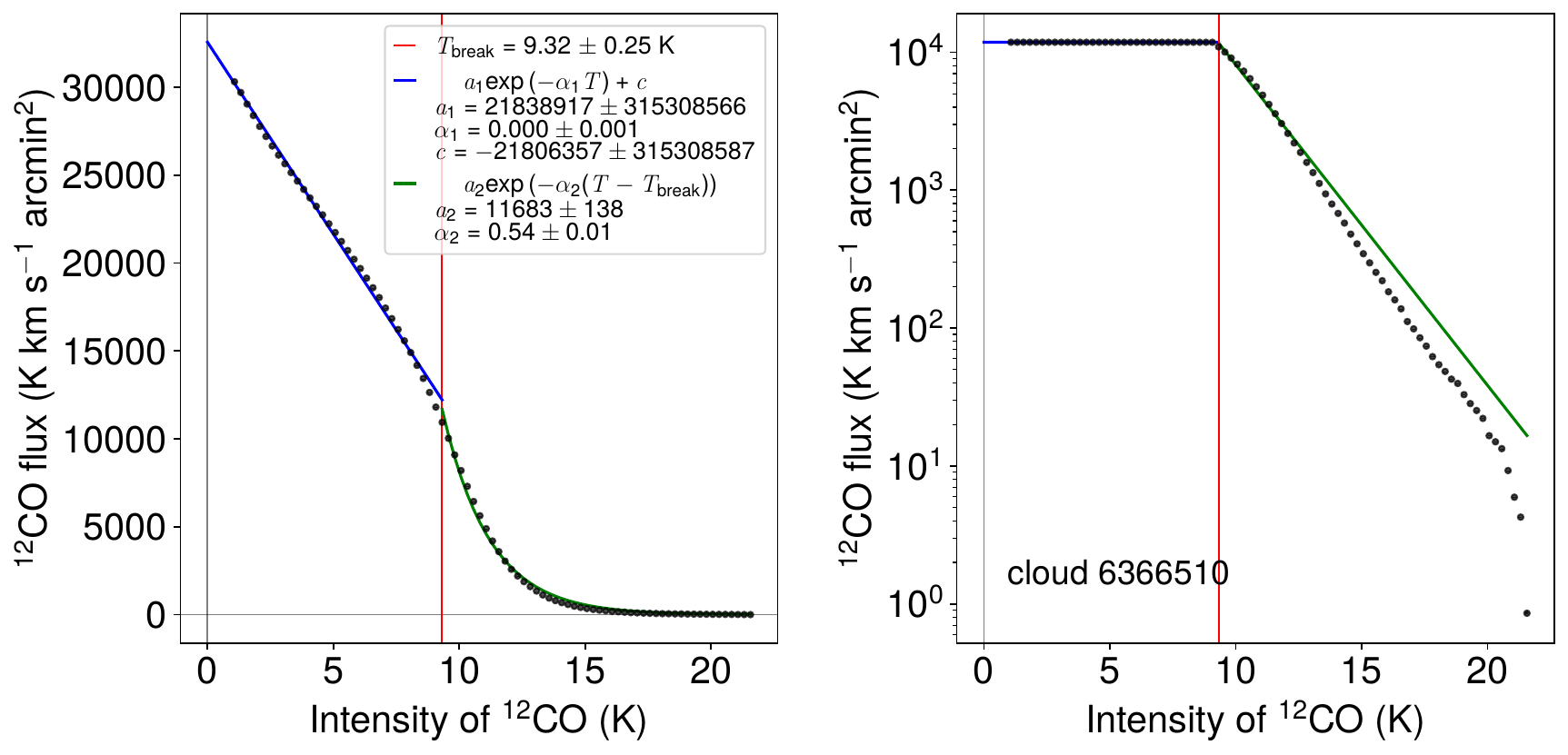}{0.33\textwidth}{(f) Flux-intensity relation of  cloud 6366510.}  } 
\caption{Integrated \cofs\ intensity images and segmented flux-intensity relations of three double-exponential  clouds. The red contour delineates the edge of the second exponential component (integrated from \tbreak\ to peak). See Table \ref{Tab:cloudpara} for detailed parameters of molecular clouds and see Equation \ref{eq:exp2} for the definition of the piecewise function. For added clarity on exponential properties, plots in logarithmic scale are also displayed, where the constant $c$ in Equation \ref{eq:exp2} has been  subtracted from the first segment, and to maintain the continuity of the plots, a shift in  logarithmic scale (by multiplying a factor) is applied to the first segment.  \label{fig:exp2three} }
\end{figure}

A brute-force search strategy is adopted to determine \tbreak. To be specific, we test each $T$ as \tbreak\  (according to the number of parameters in Equation \ref{eq:exp2}, one highest and two lowest $T$ values are excluded), and fit the remaining five parameters in Equation \ref{eq:exp2}. The fitting residual rms is used to judge the existence of \tbreak, and if the minimum fitting residual is also a local minimum, then \tbreak\ exists. In other words, moving to the left or right of \tbreak, the residual rms increases. Clearly, determining \tbreak\ in this way requires at least six data points.  In cases where no \tbreak\ is detected, this cloud is classified as single-exponential, and Equation \ref{eq:exp} is used to fit the flux-intensity relation.

Figure \ref{fig:exp2three} demonstrates images and flux-intensity relations of three representative molecular clouds. Evidently, Equation \ref{eq:exp2} fits the flux variation quite well, and the slight deviation in proximity to \tbreak\  is likely due to the fluctuation of \tbreak\ across molecular cloud regions. The \tbreak\ contours are marked with red lines in Figure \ref{fig:exp2three}, and they are highly consistent with the bright parts of molecular clouds.

\begin{figure}[h]
\gridline{\fig{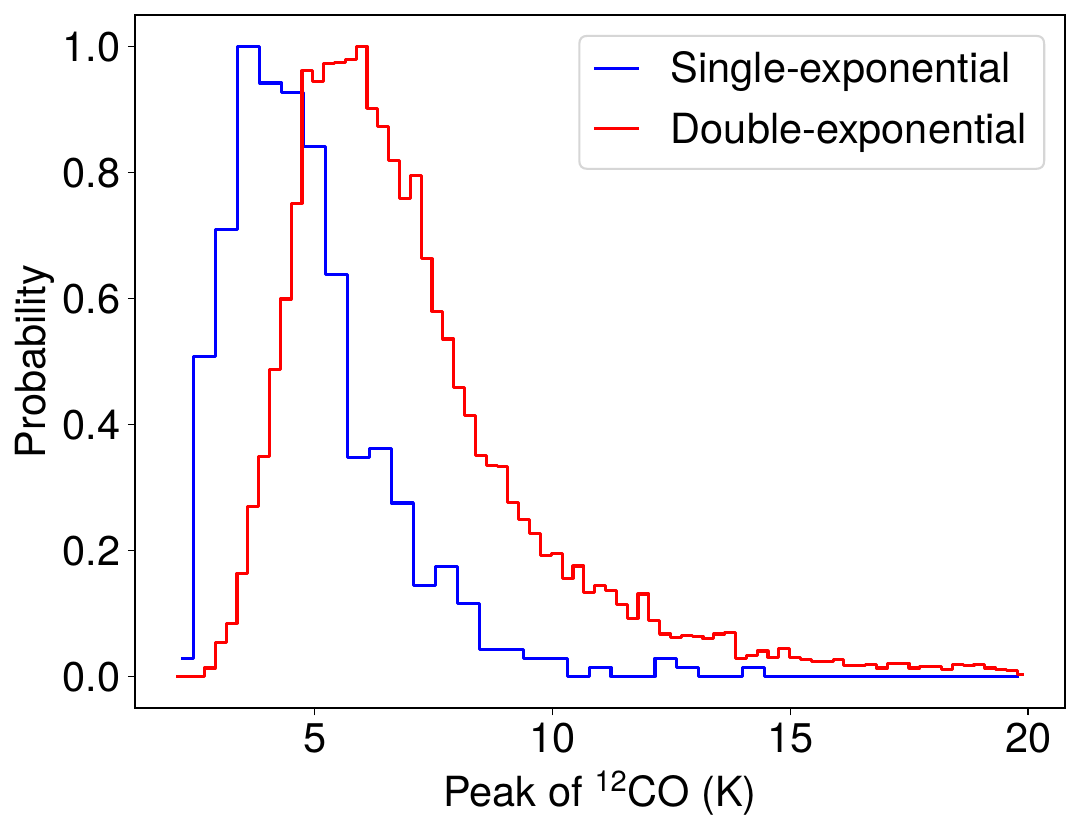}{0.3\textwidth}{(a) Normalized peak distribution.}  \fig{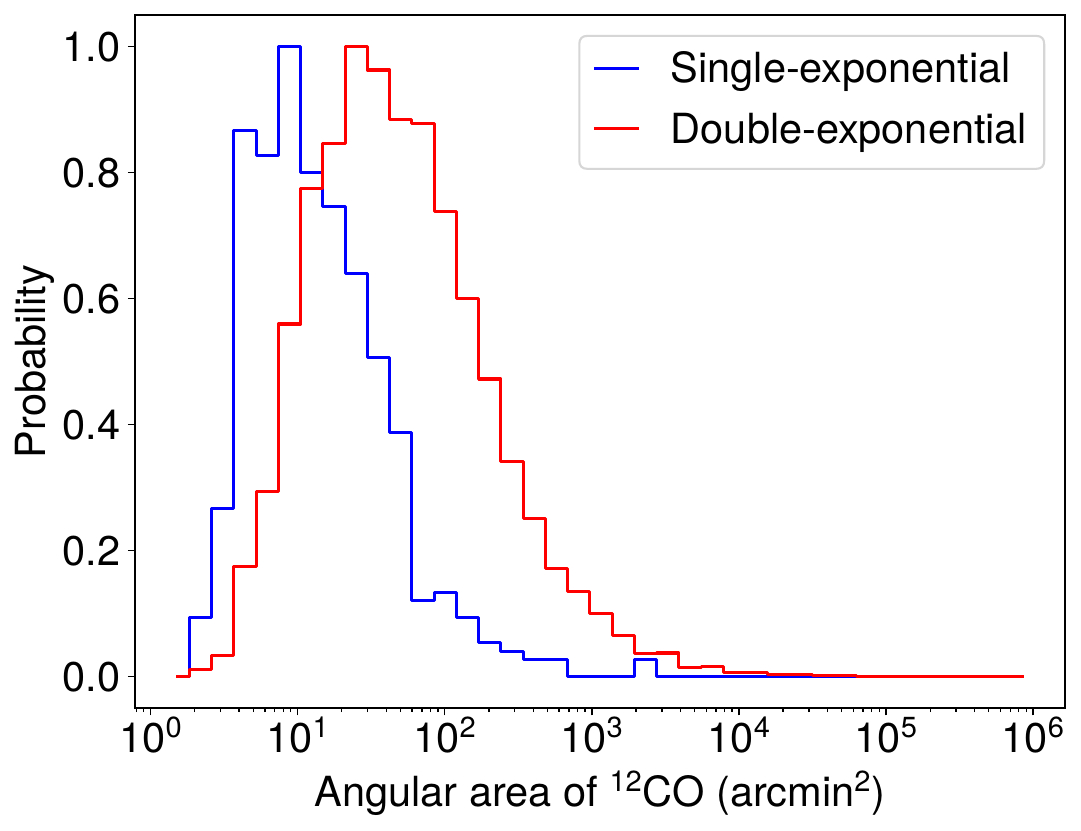}{0.3\textwidth}{(b) Normalized angular area distribution.} \fig{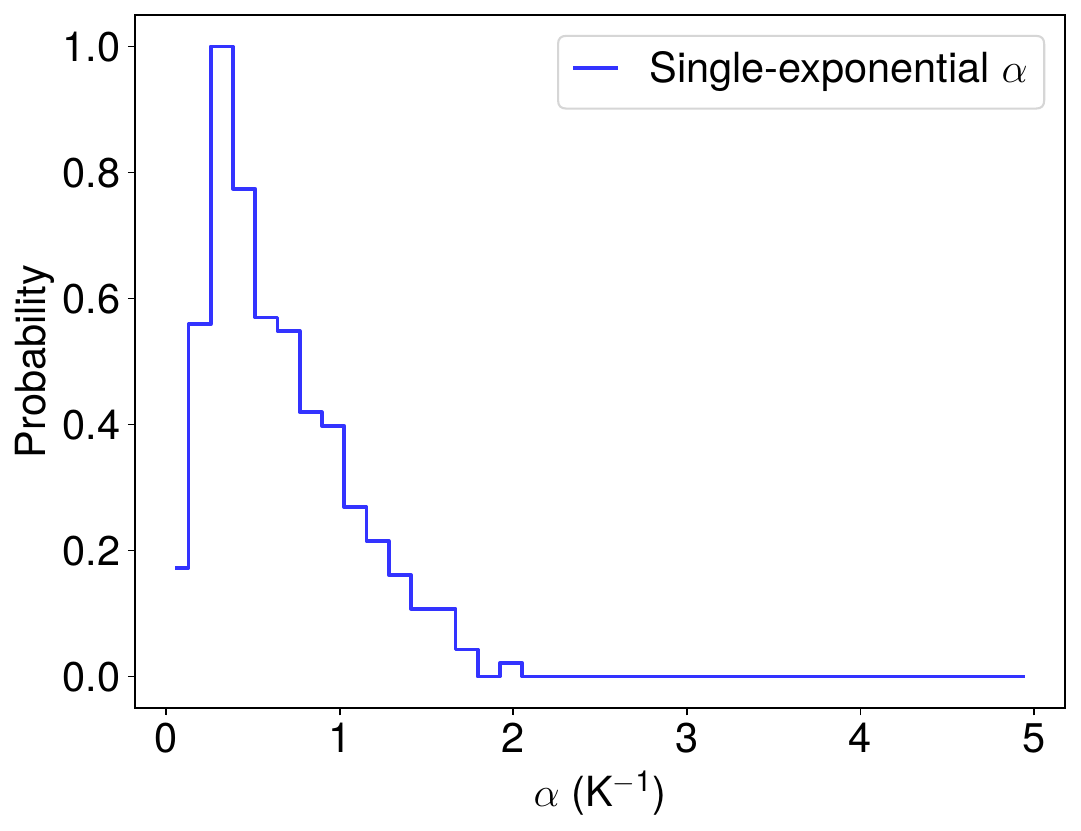}{0.3\textwidth}{(c) $\alpha$ distribution of single-exponential clouds.}     } 
\gridline{\fig{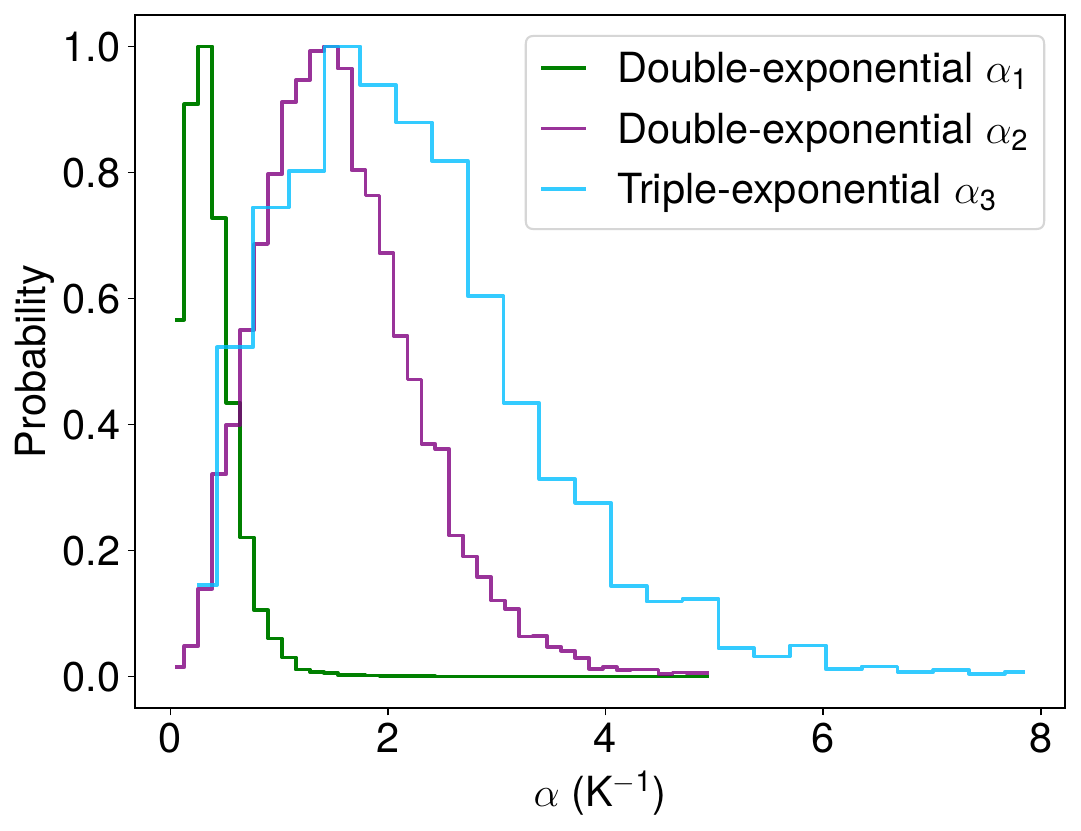}{0.3\textwidth}{(d) $\alpha$ distribution of double- and triple-exponential clouds.}  \fig{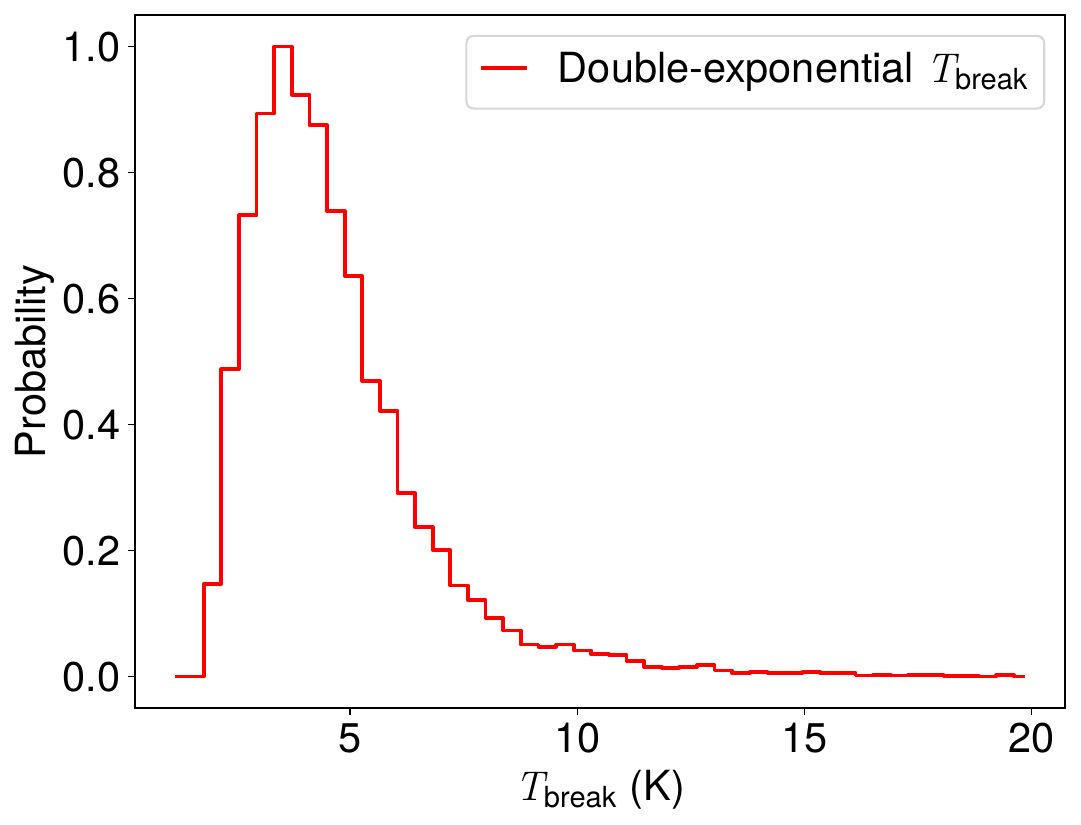}{0.3\textwidth}{(e) \tbreak\ distribution.} \fig{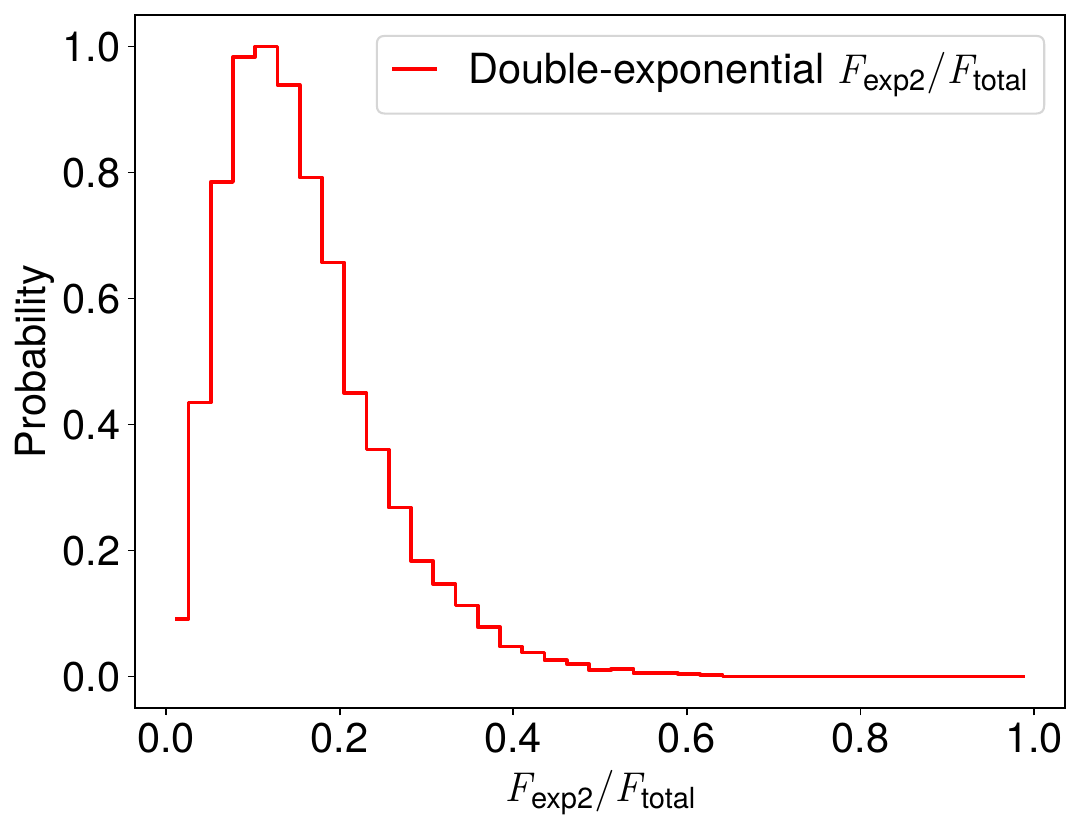}{0.3\textwidth}{(f) $F_{\rm exp2}/F_{\rm total}$ distribution.}     } 

\caption{Distribution  of the peak intensity, the angular area, $\alpha$, $\alpha_1$, $\alpha_2$, \tbreak, and $F_{\rm exp2}/F_{\rm total}$ of single- and double-exponential molecular clouds. For comparison, the steepness ($\alpha_3$) distribution of the third segment of triple-exponential \cofs\ flux-intensity relations is also plotted in panel (d). \label{fig:paradis}}  
\end{figure}

\subsection{Sampling Distribution}
\label{sec:samplingdis}

In addition to individual flux-intensity relations displayed above, we examine their distribution properties among 10866 molecular clouds in this section. Out of those clouds,  499 (4.6\%) exhibit single-exponential flux-intensity relations, and 10,367 (95.4\%) show double-exponential flux-intensity relations. With a caveat, this fraction may vary if we include those clouds whose \coss\ emission is less significant. The flux-intensity data and fitting results of flux-intensity relations of all 10,866 molecular clouds are available on the ScienceDB website (DOI:10.57760/sciencedb.17486).

Based on distributions of the peak intensity and the angular area of molecular clouds, as shown in Figure \ref{fig:paradis}, clouds with single-exponential flux-intensity relations are systematically fainter in brightness and smaller in size than those clouds with segmented flux-intensity relations.

Figure \ref{fig:paradis} also demonstrates distributions of single-exponential $\alpha$,  double-exponential $\alpha_1$, double-exponential $\alpha_2$, and \tbreak. Distributions of $\alpha$ and $\alpha_1$ are similar, peaking at about 0.3 K$^{-1}$, while $\alpha_2$ is systematically larger than $\alpha_1$ by about 1.2 K$^{-1}$. We noticed that this tendency of increasing steepness toward high intensity is maintained in triple-exponential \cofs\ flux-intensity relations, and the steepness of the third segment is systematically larger than $\alpha_2$ by about 0.6 K$^{-1}$. \tbreak\ peaks at about 3.5 K.

In addition to $\alpha$ and \tbreak, the flux fraction of the second exponential component with respect to the total observed flux (at about 1 K, 2$\sigma$) is also an important property. We define this fraction  as 
\begin{equation}
\label{eq:ratioflux}
F_{\rm exp2}/F_{\rm total}  =  \frac{ F_{\rm CO} \ \mathrm{at} \ T_{\rm break}}{ F_{\rm CO} \ \mathrm{at} \  \sim1\ \rm K},
\end{equation}
where $F_{\rm CO}$ is the CO flux of molecular clouds at a specific brightness temperature $T$. As shown in panel (f) of Figure \ref{fig:paradis}, the  $F_{\rm exp2}/F_{\rm total}$ ratio distribution reaches its maximum at approximately 0.12.

\begin{figure}[h];
\gridline{\fig{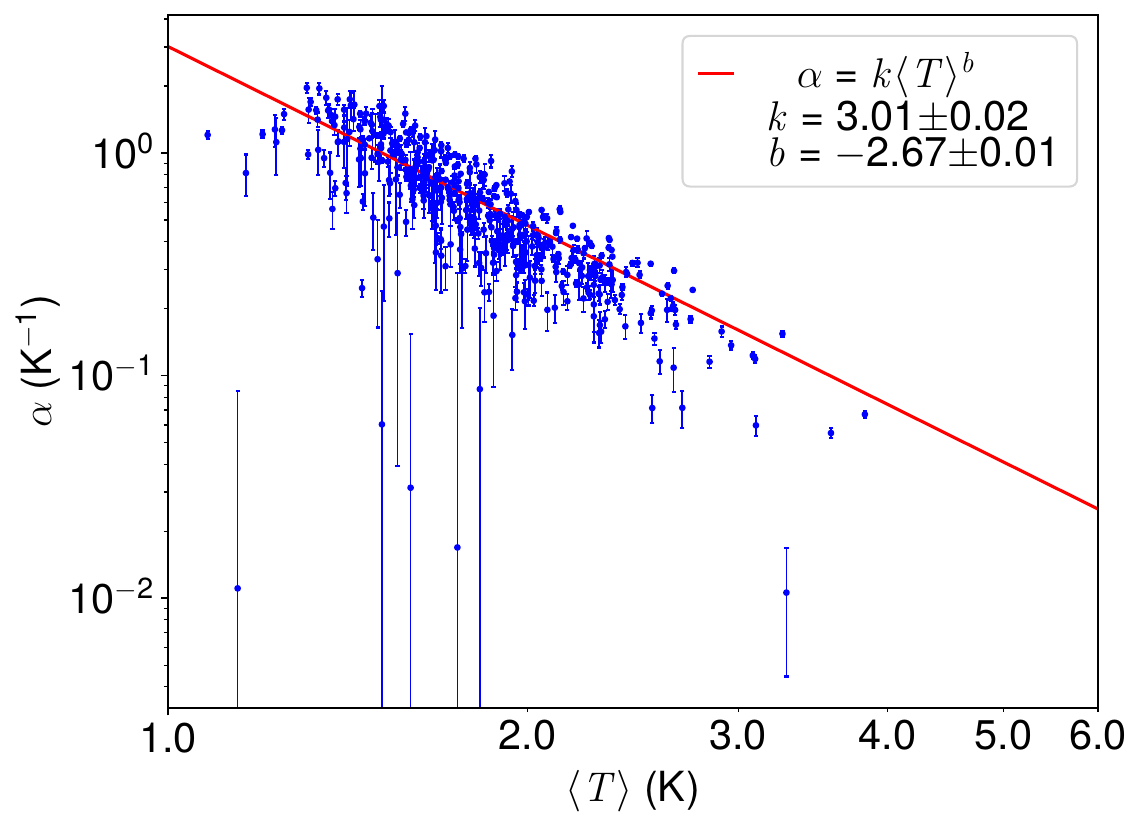}{0.4\textwidth}{(a) Correlation between $\alpha$ and \tmean.}  \fig{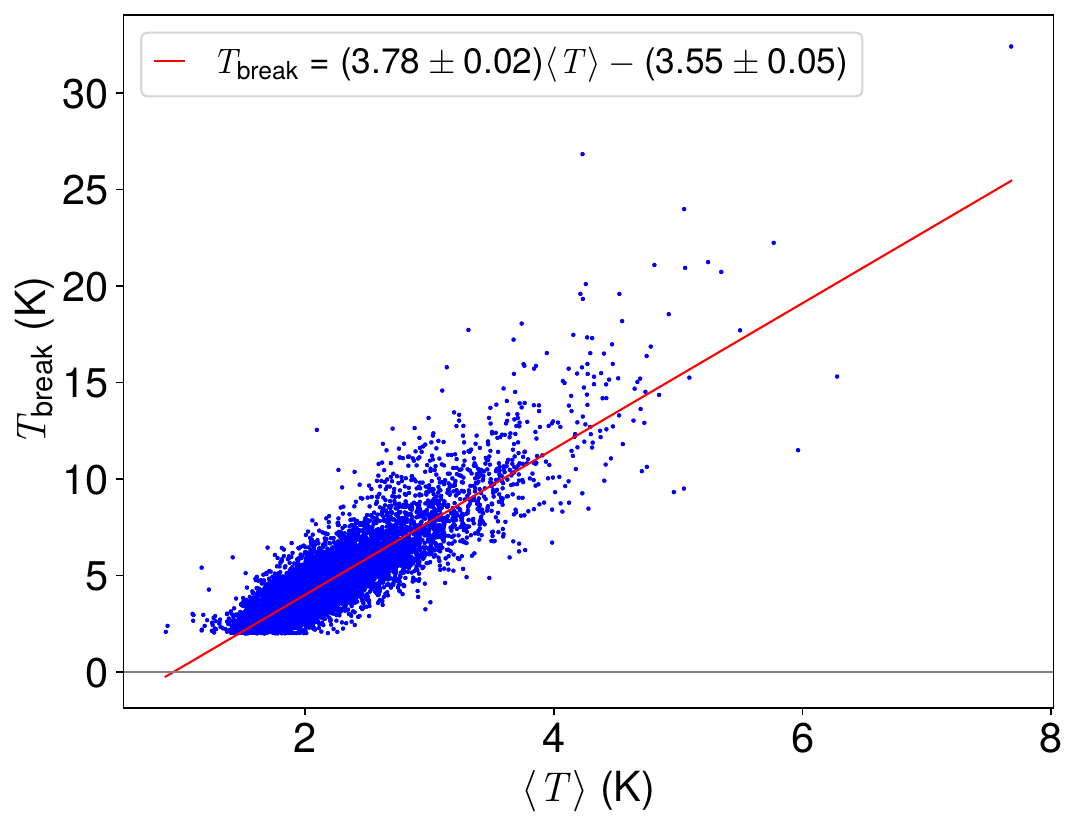}{0.4\textwidth}{(b)  Correlation between \tbreak\ and  \tmean.} } 

\gridline{ \fig{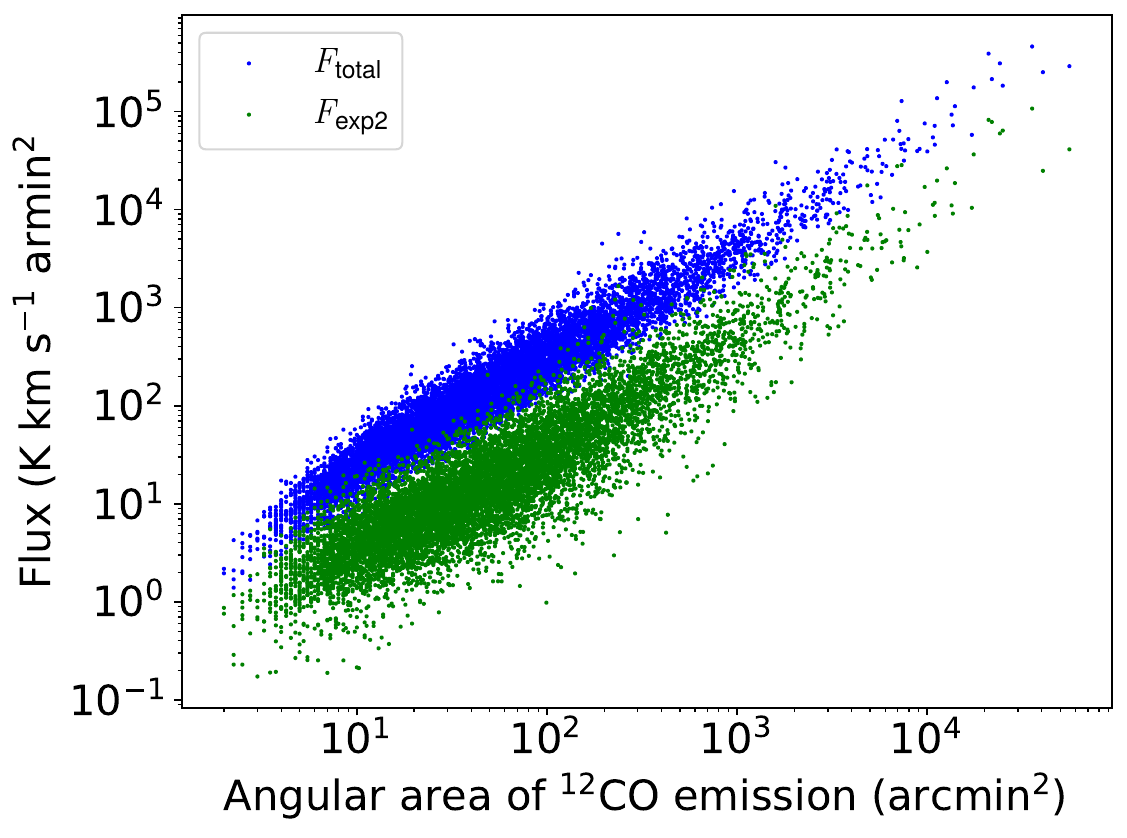}{0.4\textwidth}{(c) Correlation between flux and angular area.} \fig{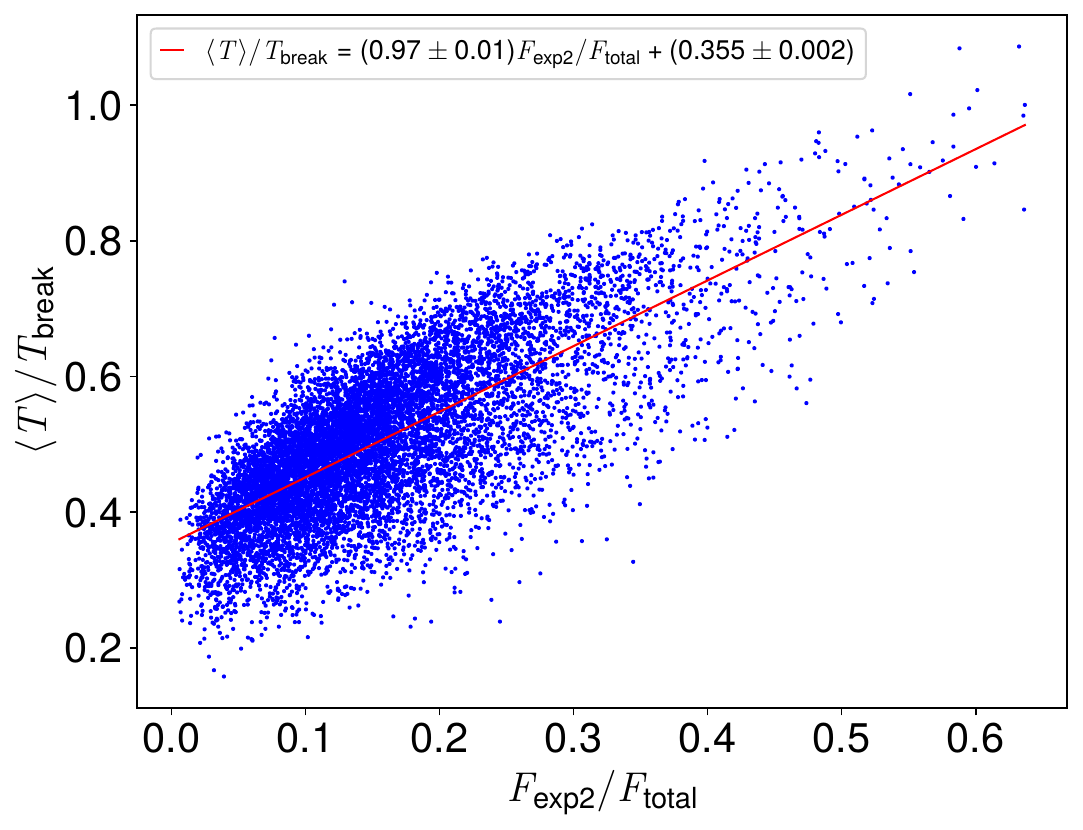}{0.4\textwidth}{(d) Correlation between $\langle T\rangle/T_{\rm break}$ and $F_{\rm exp2}/F_{\rm total}$.}    } 
\caption{Relationships between parameters of the flux-intensity relation  and cloud properties. \tmean\ is the mean brightness temperature of \cofs\ across all voxels.  $F_{\rm total}$ is the total flux of \cofs\ emission, and $F_{\rm exp2}$ is the flux of the second exponential component.  \label{fig:relations}}  
\end{figure}

\section{Discussion}

\label{sec:dis}

\subsection{Flux-Intensity Relations and Molecular Cloud Properties}


Exploring the relationships between cloud properties and the flux-intensity parameters can potentially illuminate the causes of the flux-intensity relation. 
As demonstrated in panel (a) of Figure \ref{fig:relations}, $\alpha$ is linearly correlated with \tmean\ on a logarithmic scale.  \tmean\ is the mean brightness temperature of \cofs\ across all voxels within the region of individual clouds. The single-exponential $\alpha$ ranges from 0.01 to 1.96, and clouds with low \tmean\ have higher $\alpha$ values, i.e., higher decay rates of flux. Five molecular clouds show small $\alpha$ values, corresponding to approximately linear flux-intensity relations. The sizes of those five molecular clouds are small, and their images are relatively simple. 

Surprisingly, there is a strong linear correlation between \tbreak\ and \tmean, as shown in panel (b) of Figure \ref{fig:relations}. This correlation indicates that the temperature of the second exponential component increases with the mean brightness temperature of molecular clouds. Consequently, \tmean\ is related to both the decay rate of CO flux and the break temperature of molecular clouds.

Similar to the total flux of molecular clouds, the flux of the second exponential component, $F_{\rm exp2}$,   also scales up with the angular area, as demonstrated in panel (c) of Figure \ref{fig:relations}. $F_{\rm exp2}$ occupies about 15.6\% (the mean of all double-exponential clouds) of the total flux, very close to the flux ratio (about 13\%) of  \coss/\cofs\  found by \citet{2023AJ....166..121W}.

The flux ratio, $F_{\rm exp2}/F_{\rm total}$, is approximately linearly correlated with another ratio, $\langle T\rangle/T_{\rm break}$, as demonstrated in panel (d) of Figure \ref{fig:relations}. Since the slope is close to unity, roughly,  $\langle T\rangle/T_{\rm break}$ is systematically larger than $F_{\rm exp2}/F_{\rm total}$ by about 0.355.

\subsection{\tbreak\ Contour and \coss\ Emission}

\begin{figure}[h]
\gridline{\fig{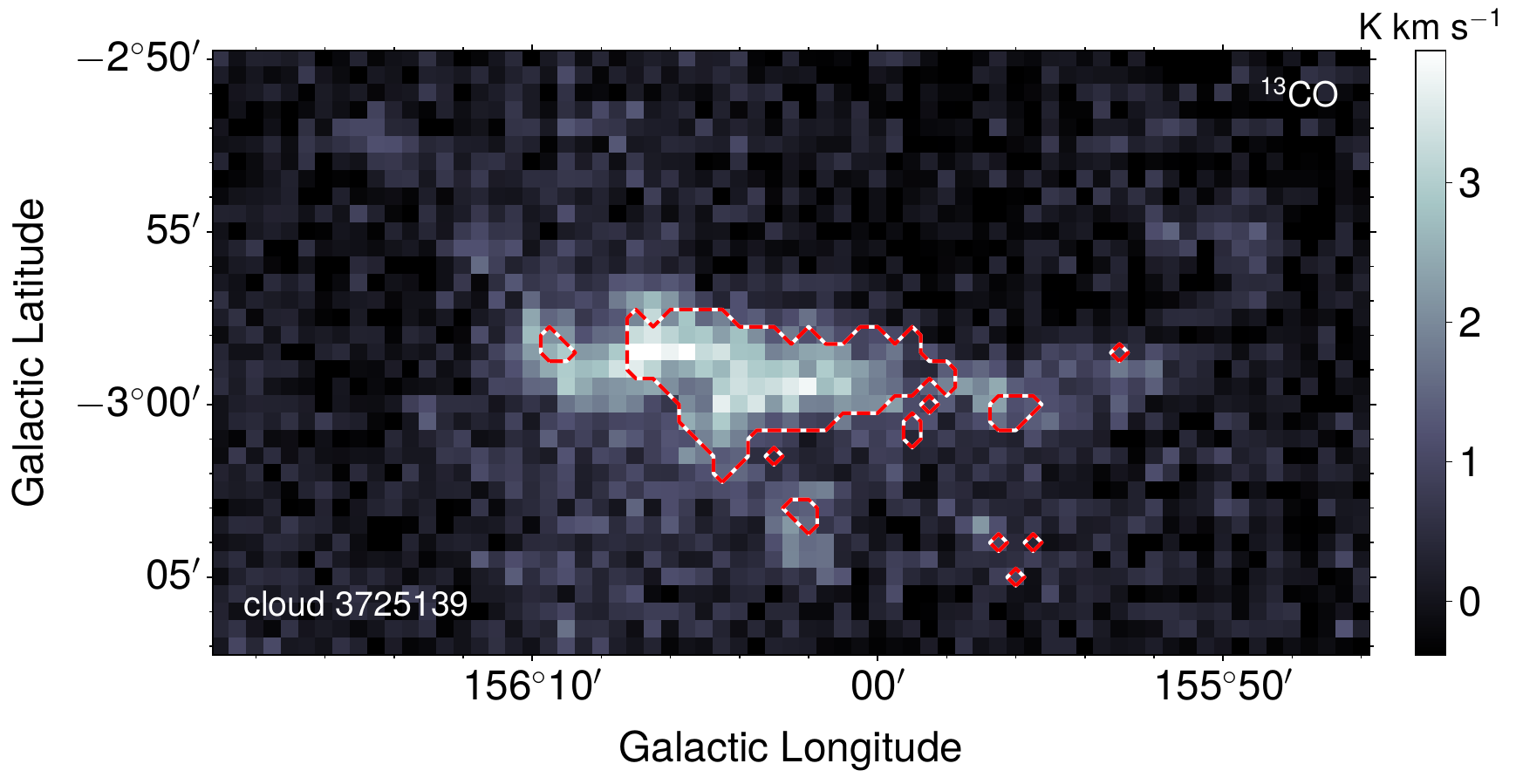}{0.35\textwidth}{(a) Cloud 3725139.} \fig{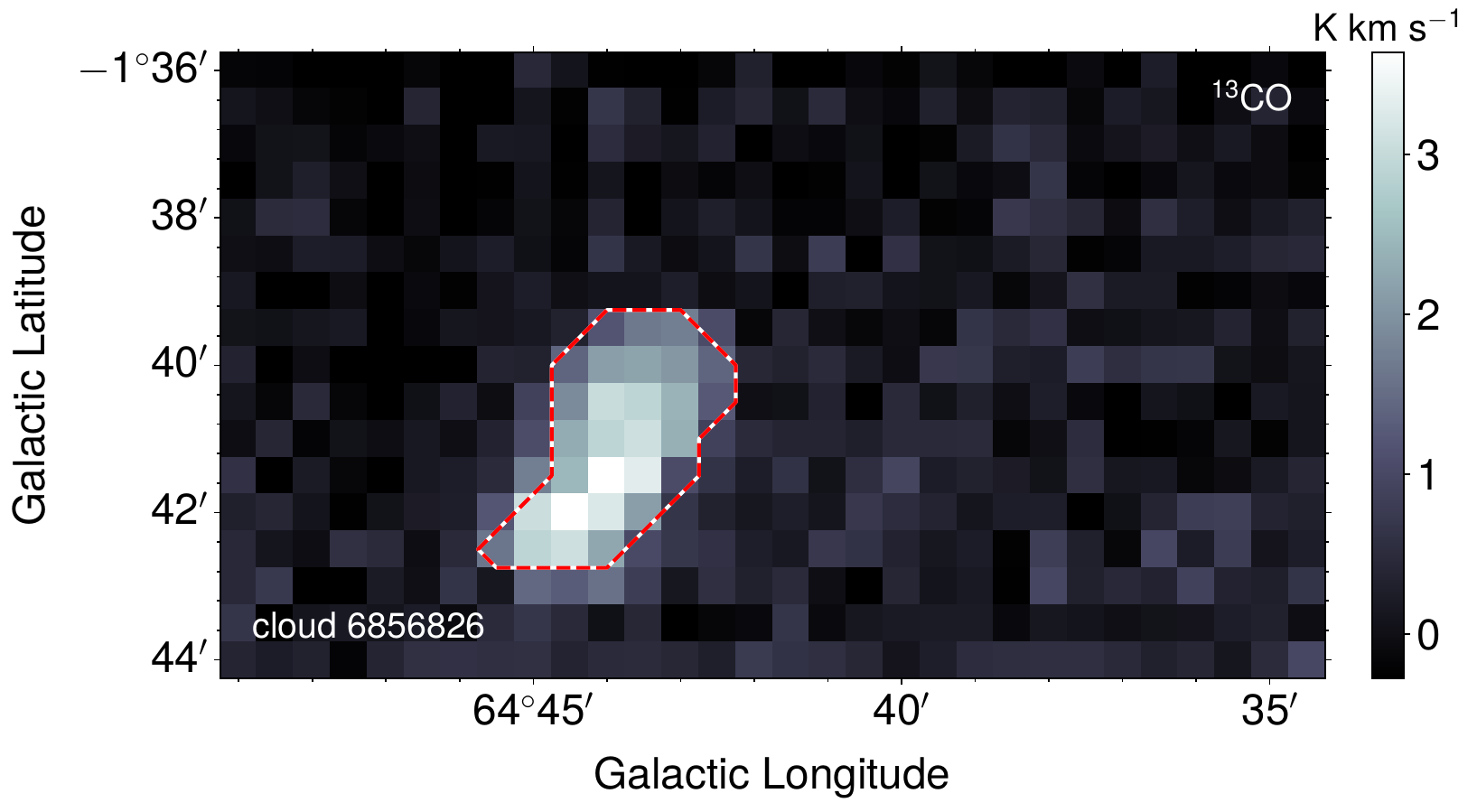}{0.35\textwidth}{(b) Cloud 6856826.}  \fig{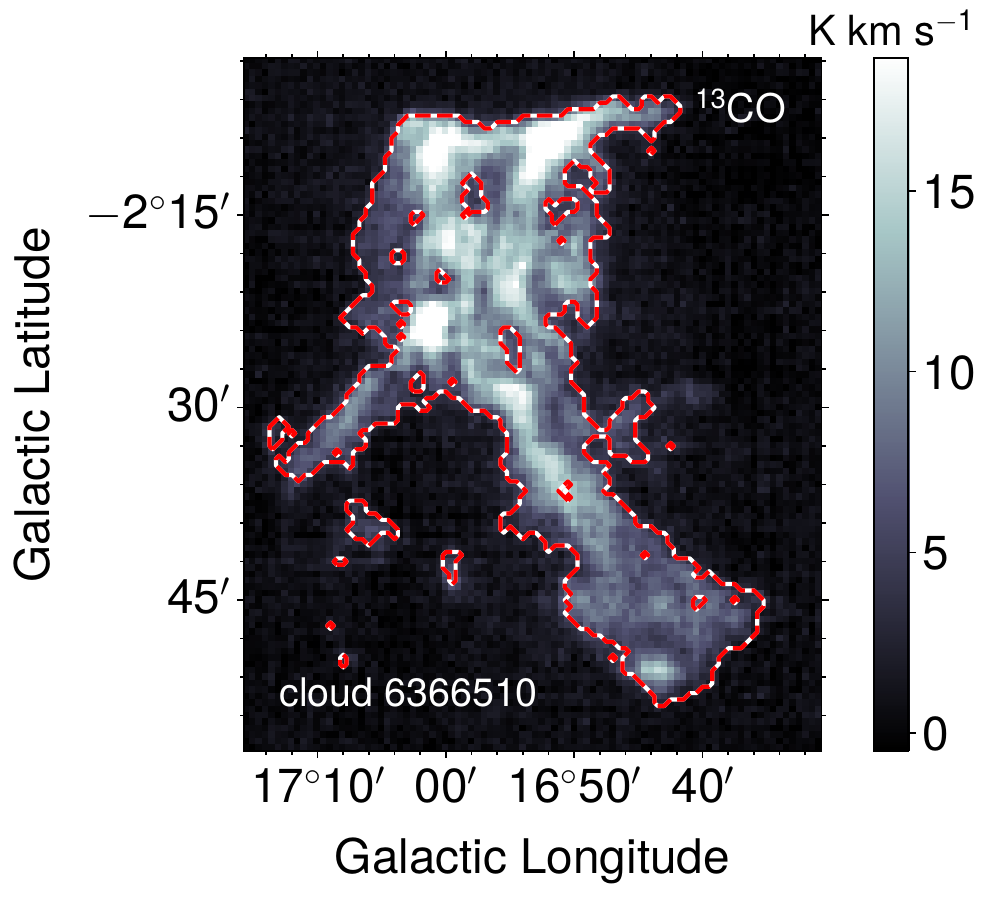}{0.22\textwidth}{(c) Cloud 6366510.}  } 
\caption{\tbreak\ contours on integrated intensity images of \coss. The \tbreak\ contour is exactly the same as that in Figure \ref{fig:exp2three}, and see Figure \ref{fig:exp2three} also for \cofs\ images and  flux-intensity relations of these  three clouds. \label{fig:imageCO13} }
\end{figure}

In addition to the correlations examined above, we also observed that the \tbreak\ contour is tightly aligned with the gas traced by \coss. This excellent correspondence between \coss\ emission and the second exponential component of \cofs\ flux-intensity relation provides valuable insight into the flux-intensity relation.

As a demonstration, we draw the \tbreak\ contour on the \coss\ map of three clouds in Figure \ref{fig:imageCO13}. The \tbreak\ contour agrees very well with the \coss\ emission at a noise rms of about 1 K \kms. This alignment indicates that the second-exponential component is tightly linked to the dense gas traced by \coss.

 Inspired by the correspondence between the \cofs\ flux-intensity relation and \coss, we inspected the \coss\ flux-intensity relation and \cots\ emission. Indeed, the edge of the second exponential component of \coss\ coincides with \cots\ emission. However, due to the low SNRs of \cots, this alignment is only seen in regions where \cots\ emission is significant.

Since \cots\ emission are imprinted on \coss\ flux-intensity relations, we may wonder what  the behavior of \cofs\ flux is at the break temperature of \coss, or equivalently, at the boundary of \cots\ emission. We found that, for \cofs\ flux-intensity relations with two break temperatures, i.e., three exponential segments, the contour of the third exponential segment, although less obvious, roughly aligns with the \cots\ emission. This confirms that the break temperature of the flux-intensity relation signifies the interface of isotopic molecular emission originating from regions where the volume densities are higher.

\begin{figure}[h]
\plotone{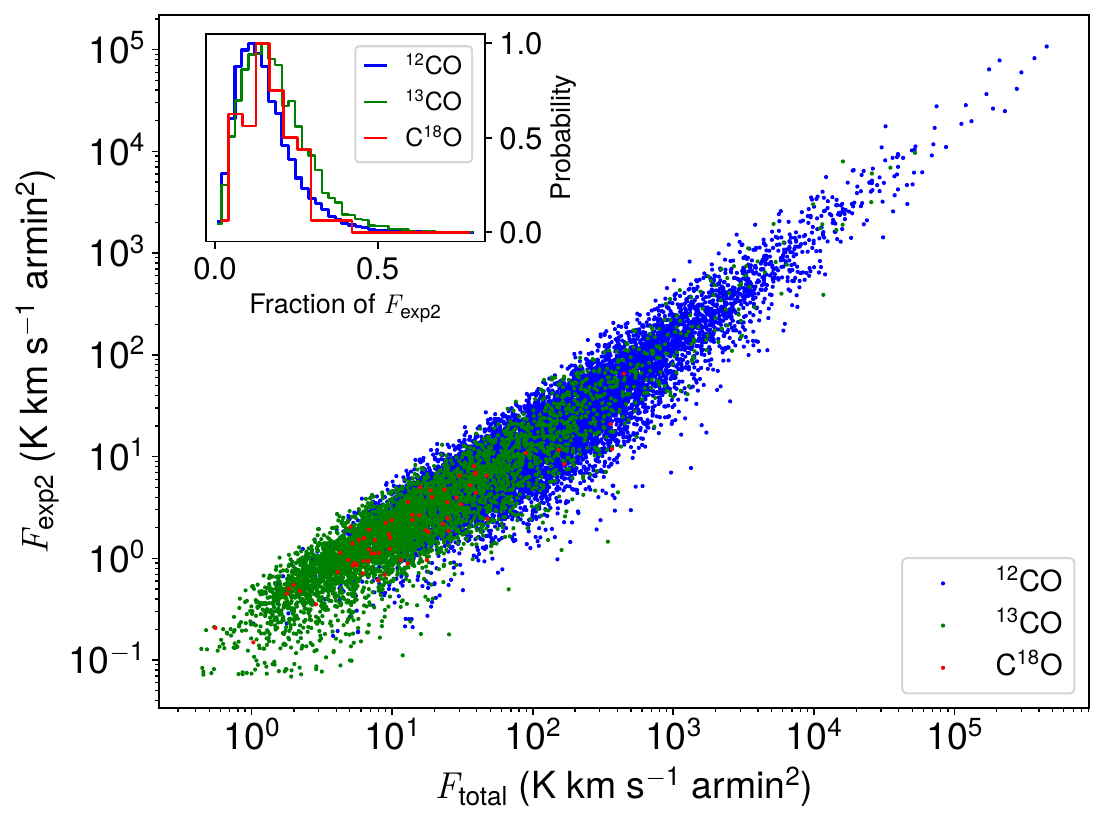}  
\caption{Correlation between the flux of the second exponential component, $F_{\rm exp2}$, and the total observed flux for \cofs\ (blue, 10,367 clouds), \coss\ (green, 4,602 clouds), and \cots\ (red, 66 clouds). Envelopes of line emission are defined using data cubes and the DBSCAN algorithm. \label{fig:exp2flux}}  
\end{figure}
\subsection{Multi-line Flux-Intensity  Relation}

In this section, we compare the flux fraction of the second exponential component among three CO lines.  Flux-intensity relations of \coss\ and \cots\ are fitted exactly the same way as \cofs.

$F_{\rm exp2}$ is roughly proportional to the total flux in logarithmic scales. This is particularly demonstrated by \cofs\ and \coss, owing to their high SNRs. As shown in Figure \ref{fig:exp2flux}, clouds with high \cots\ flux also support this scaling relationship.

As displayed in the top left corner of Figure \ref{fig:exp2flux}, distributions of the fraction of $F_{\rm exp2}$ are similar among three CO lines. The means of fractions are about 15.6\%, 18.8\%, and 16.9\%, for \cofs, \coss, and \cots, respectively.

Figure \ref{fig:exp2flux} is in line with the results of \citet{2023AJ....166..121W}, who found an excellent correlation between \cofs\ and \coss\ emissions, as demonstrated in Figure 12 therein. They derived a value of 0.13 for the ratio of \coss\ to \cofs\ emissions,  close to the mean flux fraction of the second exponential component revealed by the \cofs\ flux-intensity relation.

This strong correlation indicates that the second exponential component has physical meaning. The flux of the second exponential component scales up with the total flux of molecular clouds and shares similar fractions among molecular tracers.

\subsection{Implications of Flux-Intensity Relations}

The flux-intensity relation we found in this work exhibits two prominent features: the exponential shape and the presence of multiple exponential segments. In this section, we discuss possible implications of the shape of flux-intensity relations.

At first glance, we may suspect that the exponential shape of the flux-intensity relation is attributed to the radiative transfer process with opacity effects. Exponential segments with varying steepness represent regions with distinct optical depths. However, we found that in a simple condition, the flux-intensity relation of an isothermal sphere shows a curve with  distinct convexity, indicating that the optical depth cannot explain the shape of the flux-intensity relation. Therefore, there will be other important factors playing a  role in it.

Another key clue is the agreement between the edge of exponential components and the  isotopic molecular emission. This phenomenon indicates that segments of the flux-intensity relation correspond to regions with dissimilar compositions of materials. Based on this fact, it is reasonable to suspect that tracers of the interstellar medium, such as molecules, HI, and dust, may all exhibit flux-intensity relations with segments in their emission images, provided that those tracers can traverse across different phases of the interstellar medium. This possibility highly deserves further investigations.

 Suppose the flux-intensity relation traces the global structure of molecular clouds, then the multiple breaks in the flux-intensity relation found in this study may correspond to the historical relics of multi-episodic or hierarchical contraction. A theoretical candidate for this possibility is the Global Hierarchical Collapse (GHC) mode proposed by \citet{2019MNRAS.490.3061V}.


The flux-intensity relation fits into a broader picture of molecular clouds. The universality of the exponential shape of flux-intensity relations may be traced back to the formation process of molecular clouds.  For instance, \citet{2023ApJ...944...91Y,2023ApJ...958....7Y} proposed a picture that large-scale molecular clouds are formed by merging (coagulation). In this scenario, the flux-intensity relation constructed from a collection of individual clouds is also likely to show an exponential shape. As a preliminary proof,  we examined the total flux-intensity relation of W51, which contains many small molecular clouds \citep{2022AJ....164...55Y}, and we found that its flux-intensity relation is still exponential. Consequently, we predict that the flux-intensity relation of Giant Molecular Clouds (GMCs) in extragalaxies possesses similar shapes. However, comprehending the relationships between properties of molecular clouds, such as consistent hierarchical structures, cohesive motions of sub-structures, and universal exponential shapes of the flux-intensity relation, requires further   observations and theoretical studies.

\section{Conclusion}
\label{sec:conclusion}

We have discovered that the flux of molecular clouds above certain intensity levels, referred to as the flux-intensity relation, can be well described by exponential functions. The exponential decay rate ($\alpha$) represents the steepness of the flux-intensity relation. By examining the flux-intensity relations of 10,866 molecular clouds with high-quality images, we find that  flux-intensity relations are highly correlated with the internal structures of molecular clouds.

The primary conclusions are 
\begin{enumerate}

\item  Among molecular clouds with significant \coss\ emission, 499 (4.6\%) show single-exponential flux-intensity relations, while the remaining 95.4\% show segmented exponential forms of  flux-intensity relations. 

  \item  For \cofs, the flux fraction (about 15.6\%) and the edge of the second exponential segment are consistent with \coss\ emission. Similar relationships are also found between \coss\ and \cots. This indicates that the internal structures of molecular clouds are imprinted on CO flux-intensity relations, and segments of the flux-intensity relation correspond to regions of molecular clouds with different volume densities. 

\item  The steepness of exponential segments increases toward high-intensity regions. For double-exponential flux-intensity relations, the steepness of the second segment is systematically larger than that of the first segment by about 1.2 K$^{-1}$. The steepness further increases by about 0.6 K$^{-1}$ for the third segment, as revealed by triple-exponential  flux-intensity relations of \cofs. 

\item  The mean brightness temperature, $\langle T \rangle$, is an essential quantity of molecular clouds, and it is closely related to the decay rate of flux. The observed form is $\alpha=(3.01\pm0.02)\langle T \rangle^{-2.67\pm0.01}$.

\item  The break temperature, \tbreak, of the flux-intensity relation is also linearly correlated with $\langle T \rangle$.  The observed form is $T_{\rm break}=(3.78\pm0.02)\langle T \rangle - (3.55\pm0.05)$.

\item  The flux fraction of the second exponential component, $F_{\rm exp2}/F_{\rm total}$, is systematically smaller than the ratio of $\langle T\rangle/T_{\rm break}$ by approximately 0.355.

\end{enumerate}

\begin{acknowledgments}

This research made use of the data from the Milky Way Imaging Scroll Painting (MWISP) project, which is a multi-line survey in  \cofs/\coss/\cots\ along the northern galactic plane with the PMO-13.7m telescope. We are grateful to all the members of the MWISP working group, particularly the staff members at the PMO-13.7m telescope, for their long-term support. This work was supported by the National Natural Science Foundation of China through grant 12041305 \& 12003071. MWISP was sponsored by the National Key R\&D Program of China with grants 2023YFA1608000 \& 2017YFA0402701 and by CAS Key Research Program of Frontier Sciences with grant QYZDJ-SSW-SLH047.
 
\end{acknowledgments} 
%

\section{Software tools}

\vspace{5mm}
\facilities{PMO-13.7m telescope}


\software{astropy \citep{2013A&A...558A..33A,2018AJ....156..123A,2022ApJ...935..167A}, SciPy \citep{2020SciPy-NMeth}. }

\vspace{1mm}

\textit{Data Availability:} All parameters of 10866 molecular clouds, the flux-density data, and fitting results of the flux-density relation are  accessible at DOI:10.57760/sciencedb.17486.

 \bibliographystyle{aasjournal}
 \bibliography{refMCsample}





%

\end{document}